\documentclass[aps,prb,showpacs,superscriptaddress,twocolumn]{revtex4-1}
\usepackage{amsfonts}

\usepackage{amssymb}
\usepackage{amsmath}
\usepackage{graphicx}
\usepackage{hyperref}
\usepackage{siunitx}
\usepackage[dvipsnames]{xcolor}
\usepackage{nicematrix}
\usepackage[normalem]{ulem}
\usepackage[version=4]{mhchem}
\usepackage{float}
\usepackage{siunitx}
\usepackage[resetlabels,labeled]{multibib}
\usepackage{cancel}
\newcites{App}{App Readings}

\hypersetup{colorlinks=true,
            linkcolor=blue,
            citecolor=blue,
            urlcolor=orange}

\begin{document}

\title{The pseudogap in high-$T_c$ superconductors from SU(2) gauge symmetry\\ and dynamic correlation effects}
\author{I. A. Goremykin}
\affiliation{Center for Photonics and 2D Materials, Moscow Institute of Physics and Technology, Institutsky lane 9, Dolgoprudny
141700, Moscow region, Russia}
\author{A. A. Katanin}
\affiliation{Center for Photonics and 2D Materials, Moscow Institute of Physics and Technology, Institutsky lane 9, Dolgoprudny
141700, Moscow region, Russia}
\affiliation{M. N. Mikheev Institute of Metal Physics, Kovalevskaya Street 18,
Ekaterinburg {620219}, Russia.}

\begin{abstract}
We consider the spectral properties of the two-dimensional Hubbard model, describing the electronic properties of high-$T_c$ compounds, within the SU(2) gauge theory, which assumes the separation of electronic degrees of freedom into those of spinon and chargon subsystems. We use the dynamic mean-field theory (DMFT) approach to describe magnetic long-range order in the chargon subsystem while also treating spinon fluctuations on top of this state. We show that DMFT supplemented by long-wavelength magnetic fluctuations is essential for describing the asymmetry of the damping between the inner and outer regions of the hole pockets and the resulting formation of Fermi arcs in the underdoped regime, especially at low hole doping. The underlying hole pockets in the chargon subsystem can be associated with those observed in quantum oscillation measurements.
\end{abstract}

\maketitle

\section{Introduction}

The pseudogap phenomenon represents one of the hallmarks of the physical properties of underdoped cuprate superconductors. {The electronic spectrum in these systems has been extensively studied for a long time, both indirectly through magnetic\cite{Berthier1996Review,Asayama1996Review,Plakida},  transport \cite{Ando2004,Ando2004PhaseDiagram,Ono2007} measurements and directly by means of electron spectroscopy. In the underdoped regime, angle-resolved photoemission spectroscopy (ARPES) reveals that the low-energy electronic excitations near the Fermi level are confined to disconnected segments of the Fermi surface, commonly referred to as Fermi arcs \cite{ARPESReview_Damascelli2003,ARPES_Bi2201_DopingEvolution_Hashimoto2008,ARPES_LSCO_UnderlyingFS_Yoshida2006,ARPES_NaCCOC_Shen2005,Norman1998,Kanigel2006}. In contrast, quantum oscillation measurements performed at low temperatures and high magnetic fields provide evidence for small, closed Fermi surface pockets \cite{DoironLeyraud2007,LeBoeuf2007,DoironLeyraud2015,Yelland2008,Sebastian2011,Barisic2013,Chan2016,SebastianProust2015}. Interestingly, in some cases, the {hole pockets \cite{ARPESPockets}} or even the coexistence of Fermi arcs and hole pockets \cite{ARPES_FermiArcsPockets_Meng2009} was also observed in ARPES. 

The small Fermi surface pockets inferred from quantum oscillation measurements have been widely interpreted as arising from a reconstruction of the large Fermi surface by static density wave order, with early theoretical work demonstrating that spin-density-wave (SDW) \cite{Sebastian2008}, but also stripe \cite{Millis2007} or $d$-density-wave order \cite{Chakravarty2008} can naturally generate electron pockets consistent with experiment. Apart from that, a lot of theoretical effort was put into the description of the pseudogap phase originating from the short-range magnetic correlations. In particular, the concept of the ``nearly-antiferromagnetic'' Fermi liquid was introduced first phenomenologically \cite{Millis1990,Millis1994,Zha1996,Chubukov1996} and then put on the microscopic basis \cite{Schmalian1999,Kuchinskii1999,Onufrieva1999}. These theories consider the dynamics of spin excitations, characterized by the dynamical exponent $z=2$, which is typical for the itinerant magnetic excitations in the paramagnetic state. 

However, already in early studies, it was emphasized \cite{SokolPines1993,ChubukovPinesStojkovic1995} that the low-temperature magnetic properties of cuprates in the underdoped regime are characterized by $z=1$ excitations, which are more typical for local moment magnetic systems. The strong-coupling studies of the Hubbard model within diagrammatic quantum Monte Carlo (QMC) \cite{Wu2017,Sachdev2,Wu2018,Iskakov2024,Stepanov2026}, the dynamic cluster approach \cite{Sachdev2,Wu2018,TremblayKyungSenechal,TremblayKyungSenechal,Gull,Gunnarson,Krien2025}, and diagrammatic extensions of dynamical mean field theory (DMFT) \cite{Krien2022,Kugler2025} show pseudogap formation, induced by the delicate interplay of strong local and magnetic correlations. To describe the respective local magnetic moment degrees of freedom, the explicit separation of charge and spin degrees of freedom, whose dynamics is responsible for the characteristic features of the pseudogap regime, looks promising. This can be implemented in several ways, including the $t$–$J$ model, 
(see, e.g., Refs. \onlinecite{tJ1,tJ2,tJ3,Milstein2008})
and more recently proposed SU(2) gauge field approach, assuming the fractionalization of physical electrons into chargons coupled to an SU(2) gauge field.
The latter approach has been employed both in phenomenological \cite{NikolaenkoSachdev2023,ZhangSachdev2020,SachdevGaugeTheory2019,SachdevBergChatterjeeSchattner2016,Sachdev} and microscopic-model-based studies \cite{QiSachdev2010,SachdevHiggs1,ChatterjeeSachdevEberlein2017,ChatterjeeSachdevScheurer2017,VilardiBonetti2025, MullerGroelingBonettiForniMetzner2026, BonettiNew, ForniBonettiMullerGroelingVilardiMetzner2026, Wu2018, BonettiMetzner, Sachdev2, OurSecond, OurThird,StepanovBrenerHarkovKatsnelsonLichtenstein2022}. 

In the present paper, we study the spectral properties of cuprates within the
SU(2) gauge theory, which considers separation of electronic degrees of freedom into those of spinon and chargon subsystems, and assumes a long-range order within the chargon subsystem. {Previously, mainly the mean field treatment of chargon subsystem was considered \cite{ChatterjeeSachdevScheurer2017,VilardiBonetti2025,MullerGroelingBonettiForniMetzner2026,BonettiNew,ForniBonettiMullerGroelingVilardiMetzner2026,Wu2018,BonettiMetzner,Sachdev2,OurThird}.}
To account for dynamical correlation effects associated with the formation of well-defined local magnetic moments, the DMFT treatment of spin symmetry breaking in the chargon subsystem \cite{OurFirst,OurSecond,DMFT_Incomm1} seems more preferable. {Compared to the static mean-field description, this approach improves the applicability of the non-linear sigma model (NLSM) for describing long-wavelength fluctuations of the magnetic moments and enables one to capture effects related to quasiparticle damping.} 

Consideration of the long-range order in the chargon sector is sufficient for the formation of hole pockets and for the emergence of a pseudogap, yielding, in particular, a suppression of spectral weight in the antinodal directions. However, already at sufficiently small doping, it produces electron pockets, in contrast to the Fermi arcs typically observed in ARPES studies. To explain this change of the Fermi surface, apart from the DMFT treatment of long-range order in the chargon subsystem, we account (in contrast to some of the previous studies, e.g., Ref. \onlinecite{DMFT_Incomm1}) for the long-range fluctuations of the spin background, described by the spinon subsystem. 

By calculating the physical electron spectral functions under the assumption of fractionalization of electrons into chargons and spinons, 
we show that the formation of Fermi arcs is associated, first, with a strong anisotropy of the effective hole mass within the hole pocket and, second, with the transfer of spectral weight by magnetic fluctuations from deeper-lying electronic states, whose spectral weight differs substantially in the directions toward $\Gamma$ and $M$ points of the Brillouin zone.
{We show that DMFT supplemented by long-wavelength magnetic fluctuations is essential for describing the asymmetry in the damping between the inner and outer regions and the resulting asymmetric formation of Fermi arcs in the underdoped regime, especially at low hole doping.}

The plan of the paper is as follows. In Sec.~\ref{model_and_method_sec} we introduce the model and discuss both the chargon and spinon sectors, as well as their interplay in the properties of physical electrons. In Sec.~\ref{results_sec} we present the results obtained for the self-energies and spectral densities of chargons and electrons, as well as their Fermi surfaces. In Sec.~\ref{conclustion_sec} we present our conclusions. 

\section{Model and method}
\label{model_and_method_sec}

We consider the one-band Hubbard model 
\begin{equation}
    H =
    - \sum_{
    {i, j, \sigma
    }} 
t_{ij}{c_{i\sigma}^\dagger c_{j\sigma}}+U\sum_i{n_{i\uparrow}n_{i\downarrow}},
\label{H}
\end{equation}
with hopping $t_{ij}=t$ between nearest neighbors (which is used as a unit of energy) and $t_{ij}=-t'$ for next-nearest neighbors.

We use the fractionalized representation of electrons within the SU(2) gauge-theory framework~\cite{Schulz,Weng,Dupuis,Dupuis1,Sachdev,Wu2018,BonettiMetzner,OurSecond}. Within this approach, electrons are assumed to be separated into internal fermionic degrees of freedom, referred to as chargons (with corresponding operators $\psi,\psi^\dagger$), and a local SU(2) rotation field $R_x$, according to
\begin{equation}
\begin{split}
c_x &= \mathcal{R}_x \psi_x,\\
c^+_x &= \psi^+_x \mathcal{R}^+_x,
\end{split}
\label{c_and_psi_relation}
\end{equation}
where $\mathcal{R}_x^\dagger \mathcal{R}_x = \mathbf{1}$. The approach under consideration is based on the fact that the physical state of $c$-electron system is such that it can be mapped through the representation (\ref{c_and_psi_relation}) onto the state of chargons with long-ranged magnetic order and small fluctuations of the $R_x$ field, restoring the physical SU(2) symmetry. The dynamics of the latter is governed by the effective action of the form \cite{Dupuis,Dupuis1,BonettiMetzner,OurSecond,OurThird}
\begin{equation}
    S^{\text{eff}} [A] 
= \frac{1}{2} \sum_{x,x'} \sum_{\mu,\nu} \sum_{a,b = x,y,z} \mathcal{J}^{ab}_{\mu,x; \nu,x'} A^{a}_{\mu,x} A^{b}_{\nu,x'}
\label{A_action}
\end{equation}
with the gauge fields 
\begin{align}
    & A_{\mu x}=\mathrm{i} \mathcal{R}_{x}^{+}\partial _{\mu }\mathcal{R}_x= 
\frac{1}{2}\sum_{a=x,y,z} A_{\mu,x}^a {\sigma^a},
\end{align}
and $\mathcal{J}^{ab}_{\mu,x; \nu,x'}$ are the spin stiffnesses to be determined from a microscopic theory. In the following, we consider the chargon and spinon subsystems separately.

\subsection{Chargon sector}
\label{chargons_subsection}

\subsubsection{DMFT equations}

The chargon subsystem is described by the Hubbard model~(\ref{H}), in which the operators of physical electrons are replaced by chargon operators, $c_i \to \psi_i$; see, e.g., Refs. \onlinecite{BonettiMetzner,OurFirst,OurSecond}. {To obtain spin-dependent chargon self-energies, which correspond to the ordered states of the chargon system,  we employ dynamical mean-field theory (DMFT). Specifically, we follow the formalism described in detail in Refs.~\onlinecite{DMFT_Incomm_Licht,DMFT_Incomm,OurFirst,DMFT_Incomm1}, where antiferromagnetic states are treated by passing to a local reference frame in which the spin quantization axis at every lattice site is aligned with the local direction of magnetization. In this local frame, the standard DMFT self-consistency condition is written in terms of the lattice Green's functions of chargons in the local coordinate frame \cite{OurFirst}
\begin{align}
G^{\sigma}_{\mathbf{k},i\nu} &= -\int\limits_0^\beta d\tau \langle d_{\mathbf{k}\sigma}(\tau)d_{\mathbf{k} \sigma}^{+}(0) \rangle e^{i\nu_n \tau}\notag\\
&=
    \frac{\phi_{\nu} - \sigma \Delta_{i\nu}-(\epsilon_{\mathbf{k+Q}/2}+\epsilon_{\mathbf{k-Q}/2})/2}{(\phi_{\nu}-\epsilon_{\mathbf{k-Q/}2})(\phi_{\nu}-\epsilon_{\mathbf{k+Q/}2})-\Delta_{i\nu}^2},
\end{align}
where $\phi_{\nu} = i\nu + \mu - S_{i\nu}${ and $d,d^+$ are operators of chargons in the local reference frame \cite{OurFirst}},
\begin{align}
    S_{i\nu} &= \frac{\Sigma^{\uparrow}_{i\nu} + \Sigma^{\downarrow}_{i\nu}}{2}, \\
    \Delta_{i\nu} &= \frac{ \Sigma^\uparrow_{i\nu} - \Sigma^{\downarrow}_{i\nu} }{2},
\end{align}
{$\Sigma^{\sigma}_{i\nu}$ are the self-energies in the local reference frame}.
The local part of the lattice Green's function is assumed to be equal to the impurity Green's function: $\sum_{\mathbf k} G^{\sigma}_{\mathbf{k},i\nu}=(\zeta^\sigma_{i\nu_n}-\Sigma^\sigma_{i\nu_n})^{-1}$. In turn, the self-energy $\Sigma^\sigma_{i\nu}$ is determined from the bath Green's functions $\zeta^\sigma_{i\nu}$ by solving the respective Anderson impurity problem. }

\subsubsection{Green's function and the self-energy in the global reference frame}

For a general case of incommensurate magnetic order {at fixed $\mathcal{R}$ one can obtain} Green's function of chargons in the global reference frame $G_{\mathbf{k},i\nu}^{\text{ch}} = -\langle T {\psi}_{\mathbf{k}\sigma} (\tau) {\psi}_{\mathbf{k}\sigma}^{+}(0)\rangle_{i\nu}$ as a linear combination of Green's functions for two types of single-particle excitations \cite{OurFirst}
\begin{equation}
    G_{\mathbf{k},i\nu}^{\text{ch}}
    = \frac{1}{2} \sum_{\alpha=\pm} \mathcal{G}^{(\alpha)}_{\mathbf{k}i\nu},
    \label{G_two_types_of_excitations}
\end{equation}
where
\begin{equation}
    \mathcal{G}^{(\alpha)}_{\mathbf{k}i\nu} = \frac{1}{i\nu + \mu - \epsilon_{\mathbf{k}} - \Sigma^{(\alpha)}_{\mathbf{k},i\nu}}
\end{equation}
are the Green's functions in the local reference frame, containing the nonlocal self-energy
\begin{equation}
       \Sigma^{(\alpha)}_{\mathbf{k},i\nu} = 
S_{i\nu}  + 
\frac{\Delta_{i\nu}^2}{i\nu + \mu - S_{i\nu} - \varepsilon_{\mathbf{k} - \alpha \mathbf{Q}}}.
\label{main_general_SE}
\end{equation}

The self-energy of chargons in the global reference frame is defined as
\begin{equation}
    (G^{\mathrm {ch}} _{\mathbf{k},i\nu})^{-1} = G^{(0) \, -1}_{\mathbf{k},i\nu} - \Sigma_{\mathbf{k},i\nu}^{\mathrm{ch}}
    \label{self_energy_definition}
\end{equation}
with the free fermions' Green's function in the global reference    $G^{(0)}_{\mathbf{k},i\nu} = {1}/({i \nu + \mu -\epsilon_{\mathbf{k}}})$. In the commensurate case $\mathbf{Q} \sim -\mathbf{Q}$ (which we only consider in the following) two types of excitations (\ref{G_two_types_of_excitations}) become indistinguishable, both self-energies $\Sigma^{(\pm)}_{\mathbf{k},i\nu}$  coincide  with the chargon's self energy $\Sigma_{\mathbf{k},i\nu}^{\mathrm{ch}} = \Sigma^{(+)}_{\mathbf{k},i\nu} = \Sigma^{(-)}_{\mathbf{k},i\nu}$. 

Let us assume that the self-energy in the local reference frame demonstrates the Fermi-liquid behavior at small imaginary frequencies,
\begin{equation}
    \Sigma^\sigma_{i\nu} = \Sigma_0^\sigma + i \nu \Sigma_1^\sigma + O(\nu^2).
\end{equation}
In this case in the leading order
\begin{equation}
    S_{i\nu} =
S_0 + i\nu S_1 + O(\nu^2)
\end{equation}
and
\begin{equation}
 \Delta_\nu^2
= A_0+i
\nu A_1 
+ O(\nu^2),
\end{equation}
where we have defined the following set of low-frequency expansion coefficients
\begin{align}
S_m &= \frac{1}{2}(\Sigma_m^{\uparrow} + \Sigma_m^{\downarrow}),
\\
A_0 &= \frac{1}{4}\left( {\Sigma_0^{\uparrow} - \Sigma_0^{\downarrow}} \right)^2, \\
A_1 &= \frac{1}{2} \left({\Sigma_0^{\uparrow} - \Sigma_0^{\downarrow}}\right)\left( {\Sigma_1^{\uparrow} - \Sigma_1^{\downarrow}}\right).
\end{align}

From this we arrive at the following expression for the imaginary part of the self energy
\begin{multline}
    \text{Im} \, \Sigma^{\text{ch}}_{\mathbf{k},i\nu} = \\ = \left[ S_1 + \frac{A_i \left(\mu - \epsilon_{\mathbf{k}+\mathbf{Q}} - S_0\right) - A_0\left(1 - S_1\right)}{\left(\mu - \epsilon_{\mathbf{k}+\mathbf{Q}} - S_0\right)^2 + \left(1 - S_1\right)^2 \nu^2} \right] \nu.
\end{multline}
One can see that at the points of the Brillouin zone which fulfill
\begin{equation}
    \mu - \epsilon_{\mathbf{k} + \mathbf{Q}} - S_0 = 0,
    \label{simple_pole_equation}
\end{equation}
the Fermi-liquid behavior of the chargon self-energy is completely destroyed and
\begin{equation}
    \text{Im} \, \Sigma^{\text{ch}}_{\mathbf{k},i\nu} \Big\rvert_{\mu - \varepsilon_{\mathbf{k}} - S_0 = 0} \overset{\nu \to 0}{\sim} -\frac{A_0}{(1 - S_1)} \frac{1}{\nu}. 
\end{equation}

More generally, the symmetry breaking in the local reference frame results in singular contribution to the self-energy in the global reference frame represented by the second term in Eq. (\ref{main_general_SE}). It can be seen that it has poles at the momenta, which fulfill (cf. Eq. (\ref{simple_pole_equation})) 
\begin{equation}   
    \nu + \mu - \epsilon_{\mathbf{k} + \mathbf{Q}} - {\rm Re} S_{\nu} 
    = 0.
    \label{self_energy_poles_equation}
\end{equation}

Therefore, the singular contribution to the self-energy gives rise to zeros of the Green's function on the surface in $(\mathbf{k},\nu)$ space defined by Eq.~(\ref{self_energy_poles_equation}). At $\nu = 0$ this corresponds directly to the Luttinger surface in momentum space {at which the real part of the Green's function changes sign} \cite{Luttinger,DzyaloshinskiiLuttinger}. The important role of this surface in the pseudogap phenomenon was also emphasized previously \cite{LuttingerSurface_Kitatani2025,Sachdev2,Wu2018,ToschiLuttinger}. Quasiparticle excitations can no longer exist on this surface, leading to a complete reconstruction of the single-particle excitation spectrum.

From the point of view of the general relation between the self-energy $\Sigma_{k}^{\sigma\sigma'} (i\nu_n)$ and the full triangular vertex function $\Gamma_{k, q}^{\sigma\sigma'}$
\begin{equation}
  \Sigma_{k}^{\sigma\sigma'} (i\nu_n) = U n' \delta_{\sigma, \sigma'}
  - \frac{T}{N} \sum_{\sigma_2} \sum_{q}
  G_{k+q}^{\sigma \sigma_2} \Gamma_{k, q}^{\sigma'\sigma_2},
\label{sigma_extention_nonlocal_main}
\end{equation}
the self-energy (\ref{main_general_SE}) can be obtained by taking the vertex $\Gamma_{k, q}^{\sigma\sigma'}$ in the form
\begin{equation}
    \Gamma_{k, q}^{\sigma\sigma'} \approx \frac{\Delta_{i\nu}^2} {T} \delta_{\mathbf{q},\mathbf{Q}}  ,\delta_{\omega_n,0} \delta_{\sigma,\sigma'}.
    \label{vertex_approximation}
\end{equation}

Thus, by considering a system of chargons and writing their self-energy in the form~(\ref{main_general_SE}), we include only static and homogeneous magnetic correlations. We now aim at going beyond this approximation and take into account the effects of long-wavelength dynamical magnetic fluctuations by considering spinons.

\subsection{Spinon sector}
In the rest of the paper we consider antiferromagnetic order characterized by magnetic wave vector $\mathbf{Q}=(\pi,\pi)$ and respective local spin stiffnesses having the form \cite{BonettiMetzner,OurSecond}
\begin{equation}
    \mathcal{J}^{ab}_{\mu,x; \nu,x'} =  
\begin{pmatrix}  
J_{\mu,x;\nu,x'} & 0 & 0 \\  
0 & J_{\mu,x;\nu,x'} & 0 \\  
0 & 0 & 0  
\end{pmatrix}.
\end{equation}
In this case the SU(2) field $\mathcal{R}_x$ can be parameterized in terms of a two-component bosonic field $z_x,z^\dagger_x$
\begin{equation}
    \mathcal{R}_x =
\begin{pmatrix}  
z_{1,x}^* & z_{2,x}^* \\  
- z_{2,x} & z_{1,x}  
\end{pmatrix}
\in SU(2),
\label{R_parametrization}
\end{equation}
which corresponds to the bosonic (spinon) degrees of freedom satisfying the local constraint
\begin{equation}
    z_x^+ z_x = z^*_{1,x} z_{1,x} + z^*_{2,x} z_{2,x} = 1.
    \label{z_constraint}
\end{equation}
For this particular choice of parametrization, the gauge field $A^{a}_{\mu,x}$ can be expressed in a convenient CP$^{N-1}$ form 
\begin{equation}
    A_{\mu,x}^a
=\mathrm{i} \Big[(\partial_\mu z_x^\dagger)\,\sigma^a\,z_x
- z_x^\dagger\,\sigma^a\,(\partial_\mu z_x)\Big]
\label{A_direct_z}.
\end{equation}
Substituting this expression into the action~(\ref{A_action}), we can then employ a saddle point approximation to study the dynamics of the field $z_x$, with the constraint $z_x^\dagger z_x = 1$ enforced on average, see the details of the derivation in Appendix \ref{z_field_mean_field_derivation}. The propagator of $z_x$-field, defined as
\begin{equation}
    D_{x,x'}^{\sigma\sigma'} = \langle z^{+\,\sigma}_x z_{x'}^{\sigma'} \rangle,
\end{equation}
is assumed to be symmetric and has a simple form
\begin{equation}
    D_{x,x'}^{\sigma\sigma'} =D_{x,x'} \delta_{\sigma\sigma'},
    \label{D_xx}
\end{equation}
with the Fourier transform of $D_{x,x'}$, given in the mean-field approximation by
\begin{equation}
    D_k = \frac{1}{2 \Delta + \Sigma_k},
    \label{z_propogator}
\end{equation}
where $\Delta$ is the magnetic energy gap,  $\Sigma_k=\tilde{\Sigma}_k - \tilde{\Sigma}_0$ depends self-consistently on $D_k$ as
\begin{multline}
    \tilde{\Sigma}_k =  \sum_{\mu,\nu}   \sum_q J_{\mu\nu,q} \sum_{\alpha = \pm} D_{k+\alpha q}\\ \times 
\Big[
(2 k_{\mu} + \alpha q_{\mu}) (2k_{\nu} + \alpha q_{\nu}) + q_{\mu} \,q_{\nu}
\Big],
\label{sigma_k_expression}
\end{multline}
and $J_{\mu\nu,q}$ is the Fourier transform of the spin stiffnesses.

Assuming that most of the spectral weight carried by $D_q$ is concentrated near $q = 0$, we may approximate $\Sigma_k$ as
\begin{equation}
    \Sigma_k = 2 \sum_{\mu,\nu} J_{\mu\nu,k} \,k_{\mu} k_{\nu},
\end{equation}
which results in the sum-rule equation in the form studied previously in Ref. \cite{OurThird}.

For simplicity, we further assume that spin stiffness is fully local in space and time ${J}_{\mu,x; \nu,x'} \approx {J}_{\mu \nu} \; \delta_{x,x'}$, as it was assumed in most of previous studies (see, e.g., Ref. \onlinecite{BonettiMetzner}). Arguments supporting this approximation for the DMFT calculations considered here can be found in Appendix~\ref{temporal_spin_stiffness_ap}. In this case even simpler equation for $\Sigma_k$ appears
\begin{equation}
    \Sigma_k = 2 \sum_{\mu,\nu} J_{\mu\nu} \,k_{\mu} k_{\nu}.
\end{equation}
For the considered case of antiferromagnetic order, $J_{\mu\nu}$ is diagonal, and the bosonic propagator can be written in the standard form
\begin{equation}
    D_{\mathbf{q},i\omega_n} = \frac{1}{2} \frac{1}{\Delta + \chi \omega_n^2 + \rho (q_x^2 + q_y^2)},
        \label{D_propogator}
\end{equation}
where we have introduced spatial spin stiffness $\rho$ and temporal spin stiffness $\chi$, which is assumed to be static.

The magnetic energy gap $\Delta$ can be determined from the sum-rule 
\begin{equation}
    2 \sum_k D_k =2 T \sum_{\mathbf{q,i\omega_n}} D_{{\mathbf q},i\omega_n}= 1,
    \label{sum_rule_equation}
\end{equation}
where factor $2$ follows from two components of full propagator (\ref{D_xx}).
With assumptions described above, the sum over frequencies can be easily carried out analytically leading to the equation
\begin{equation}
    \sum_{\mathbf{q}} \frac{1}{2 \chi \omega_{\mathbf{q}}}{\coth\frac{\omega_{\mathbf q}}{2T}} = 1.
    \label{gap_equation_after_wn_sum}
\end{equation}
with the bosonic dispersion $\omega_{\mathbf{q}}$ in the form
\begin{equation}
    \omega_{\mathbf{q}} = 
\sqrt{\omega_0^2 + v^2 q^2},
\label{spinon_dispersion}
\end{equation}
where $v = \sqrt{{\rho}/{\chi}}$ is the spin-wave velocity and $\omega_0 = \sqrt{{\Delta}/{\chi}}$.
The sum over the momentum $\mathbf{q}$ is carried out in a certain cutoff radius $\Lambda$ which if ultraviolet cutoff for the theory (\ref{A_action}).

{We determine the spatial spin stiffness $\rho$ from the $yy$ component of the spin current correlator of the chargon sector, 
$ \rho = M^{yy}_{q\rightarrow 0;xx}$, 
see Ref.~\onlinecite{OurSecond}. The static temporal spin stiffness is determined directly as $\chi ~=~ \lim\limits_{\omega \to 0} m^2 / (\chi^{yy}_{\mathbf{q} = \mathbf{Q}, \omega}  \omega^2)$ where the frequency dependence of the susceptibility on the imaginary axis is obtained by polynomial continuation.}

\subsection{Physical electrons}

From Eq. (\ref{c_and_psi_relation}) it follows that Green's functions of chargons $G$ and physical electrons $G^g$ are related by
\begin{equation}
    G_k = 2\sum_q  D_q \, \text{Tr} \, G^{\mathrm{ch}}_{k - q}.
\end{equation}
We are particularly interested in the spectral function of electrons 
\begin{multline}  
A_{\mathbf k}(\nu) = -\frac{1}{\pi}{\mathrm Im} G_k=  
\iint \frac{dq_x\,dq_y}{(2\pi)^2}\,  
\frac{1}{2\chi\,\omega_{\mathbf q}} \\  
\times\Bigg[
A^\mathrm {ch}_{\mathbf{k}-\mathbf q}\bigl(\nu-\omega_{\mathbf q}\bigr)\,  
\Bigl(n_B(\omega_{\mathbf q}) + n_F\bigl(\omega_{\mathbf q} - \nu\bigr)\Bigr)  
\\  
+ A^\mathrm {ch}_{\mathbf{k}-\mathbf q}\bigl(\nu+\omega_{\mathbf q}\bigr)\,  
\Bigl(n_B(\omega_{\mathbf q}) + n_F\bigl(\omega_{\mathbf q}+\nu\bigr)\Bigr)
\Bigg],
\label{A_g_convolution}
\end{multline}
where $A^{\rm ch}_{\mathbf{k}}(\nu)$ is the spectral function chargons and we have used the spectral function of magnetic fluctuations (\ref{D_propogator}) 
\begin{multline}
    A^b_{\mathbf{q},\omega} = -\frac{1}{\pi} \text{Im} \, D_{\mathbf{q},\omega+i 0^+} = \\
    = 
    \frac{1}{{4}\chi\,\omega_{\mathbf q}}
\left[
\delta\!\left(\omega+\omega_{\mathbf q}\right)
- \delta\!\left(\omega-\omega_{\mathbf q}\right)
\right].
\end{multline}
For simple models of the electronic dispersion, such as the hole pocket considered in Appendix~\ref{hole_pocket_analysis}, Eq.~(\ref{A_g_convolution}) allows one to obtain semi-analytical results and explain results of full the calculation.
{In general instead of well-defined chargon dispersion branches, the physical electrons exhibit a rather broad excitation continuum. At each momentum, the spectral weight is concentrated near square-root singularities. The positions of these square-root singularities define a new effective electronic dispersion. Because the spinon spectrum is gapped by $\Delta$, this effective dispersion is split by an amount $2\Delta$ in the vicinity of band extrema, in particular near the chargon hole pocket. Depending on the effective mass in the hole pocket, this splitting can either lead to a complete suppression of the spectral weight at the Fermi level or shift it in momentum space across the Brillouin zone.}

\section{Results}
\label{results_sec}

We have performed DMFT calculations for $U=5.6t$ on a square lattice with next-neighbor hopping $t'=0.3t$ for several temperatures. {Throughout the paper, we use $t$ as the unit of energy.}

Fig.~\ref{fig_spectral_functions} shows the {chargons} spectral functions for several doping levels. The significant redistribution of spectral weight is observed when the doping is increased from $x = 1\%$ to $x = 5\%$. Increasing the hole concentration reduces the spectral weight in the Hubbard bands while enhancing the weight of the quasiparticle peak. The resulting evolution of the spectral function leads to a gradual filling of the spectral gap, which we associate with the formation of a pseudogap. 

\begin{figure}[t]
    \centering
    \includegraphics{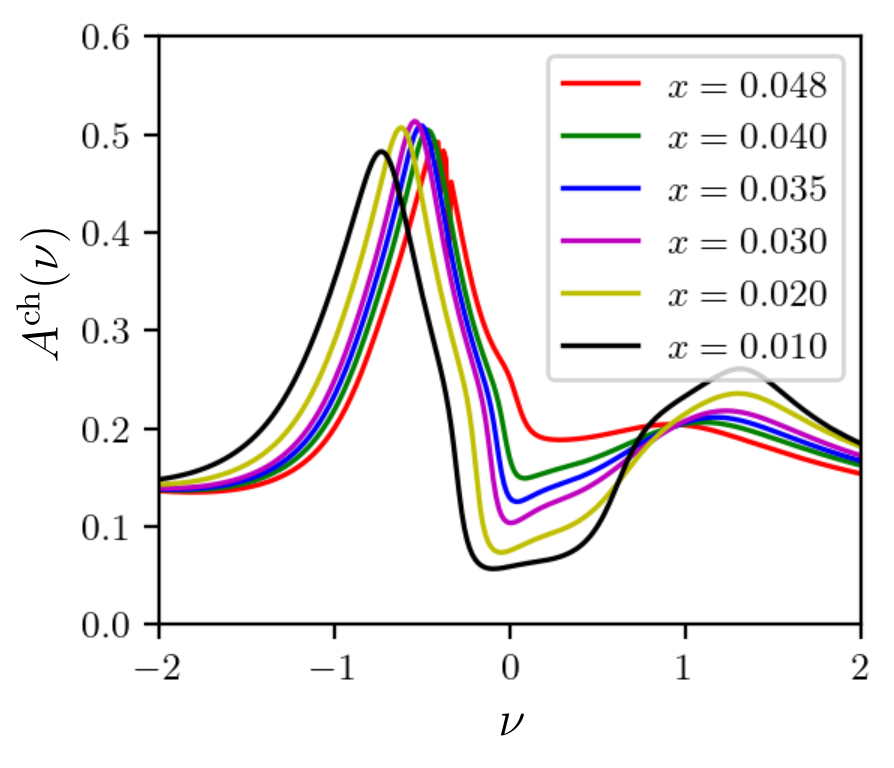}
    \caption{Spectral functions of chargons with pseudogap formation for $U=5.6t$ for various levels of hole's concentrations $x$ at $T=0.1t$.}
\label{fig_spectral_functions}
\end{figure}

\begin{figure}[b]
    \centering
    \includegraphics{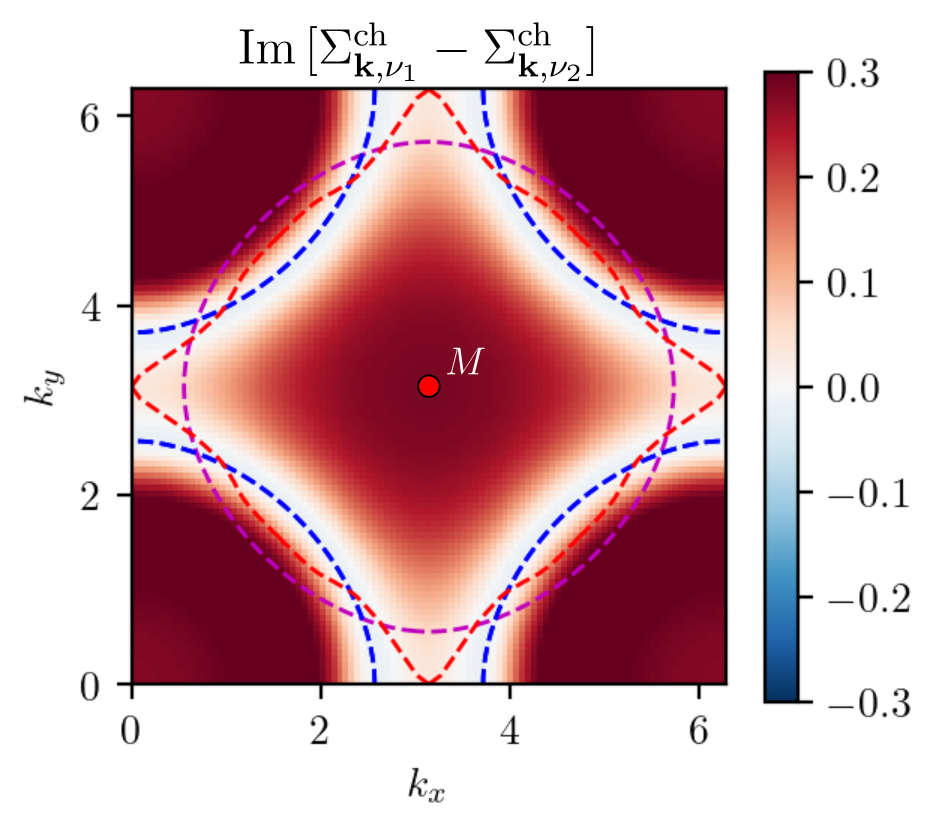}
    \caption{Imaginary part of the difference of self-energies of chansons at first and second Matsubara frequencies $\Sigma^{\text{ch}}_{\mathbf{k},\nu_1} - \Sigma^{\text{ch}}_{\mathbf{k},\nu_2}$  as a function of wave vector $\mathbf{k}$ for $x=0.04$ for $T=0.2 t$. The blue, red, and purple dotted lines denote the solution of Eq.~(\ref{self_energy_poles_equation}), the local extrema of $A^{\text{ch}}_{\mathbf{k}}(\nu=0)$, and the noninteracting Fermi surface, respectively.}
    \label{self_energy_difference_a}
\end{figure}


\begin{figure*}[t]
    \centering
\includegraphics[width=0.8\linewidth]{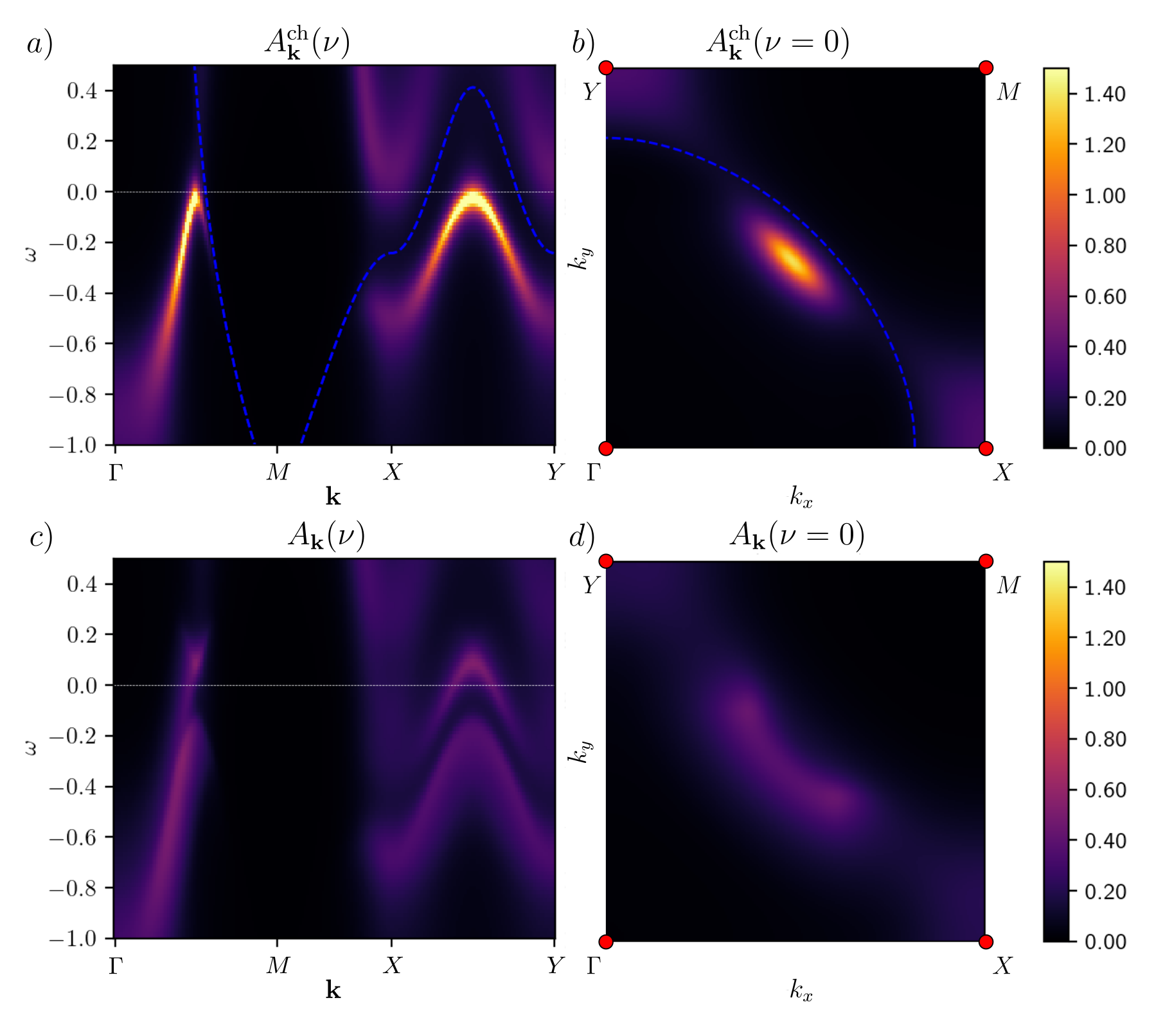}
    \caption{Spectral density of chargons (a,b) $A^{\text{ch}}(\mathbf{k},\omega)$ and physical electrons (c,d) $A(\mathbf{k},\omega)$ and for $U=5.6t$ and $x=0.040$ at $T=0.2 t$. The blue dotted line corresponds to the solution of Eq.~(\ref{self_energy_poles_equation}).
    }
\label{beta_5_spectral_results}
\end{figure*}

To investigate the developed pseudogap state, we consider in the following the state  with hole concentration $x = 4\%$ ({the so called Wu point, Ref. \onlinecite{Wu2017})}. As an indicator of the regions in $\mathbf{k}$-space where fermionic quasiparticles in the chargon subsystem are completely destroyed due to divergences of the self-energy, we plot ${\rm Im}(\Sigma^{\text{ch}}_{\mathbf{k},\nu_1} - \Sigma^{\text{ch}}_{\mathbf{k},\nu_2})$ at the first and second Matsubara frequencies $\nu_{1,2}$. The corresponding results are shown in Figs.~\ref{self_energy_difference_a} for $T=0.2t$. One can see that the non-Fermi-liquid region, where this difference is negative, is concentrated along the line defined by Eq.~(\ref{self_energy_poles_equation}).

In Figs.~\ref{beta_5_spectral_results}a,b 
we present the computed spectral functions for chargons, $A^{\text{ch}}_{\mathbf{k}}(\nu)$, together with the corresponding Fermi-level spectra, $A^{\text{ch}}_{\mathbf{k}}(\nu=0)$ at $T=0.2 t$.
One can see that  the nonlocal spectral function $A^{\text{ch}}(\mathbf{k},\omega)$ of chargons and the corresponding Fermi surface, determined by the maxima of  $A^{\text{ch}}(\mathbf{k},\omega=0)$, are strongly broadened: single-particle excitations do not form coherent states at the Fermi level, but only a diffuse feature in the vicinity of $(\pi/2,\pi/2)$. In the regions fulfilling equation (\ref{self_energy_poles_equation}) there is a strong suppression of chargon spectral weight, as expected. The local extrema of $A^{\text{ch}}_{\mathbf{k}}(\nu=0)$ (see also Fig. \ref{self_energy_difference_a}), extend well into the region associated with the Fermi-liquid behavior only along the nodal direction, thereby forming, at the present temperature, hole pockets.

We then consider the spectral functions of electrons, $A_{\mathbf{k}}(\nu)$, obtained from Eq.~(\ref{A_g_convolution}). Following Refs.~\onlinecite{OurSecond,OurThird}, we use Eq.~(\ref{gap_equation_after_wn_sum}) to determine the magnetic gap $\Delta$ and set the cutoff $\Lambda=0.5$. For $T=0.2 t$ we estimate $\rho=0.033t$, $\chi=0.120t^{-1}$, which yield $\omega_0 \approx 0.11 \, t$ and the correlation length $\xi/a=\sqrt{\rho/\Delta}\approx4.9$.
In Fig. \ref{beta_5_spectral_results} c,d) 
we present the results obtained for the electronic spectral functrions $A$ after the convolution of chargon spectral functions $A^{\text{ch}}$ with the bosonic propagator $D_{\mathbf{q},\omega}$ according to the Eq. (\ref{A_g_convolution}).
{The resulting spectral density exhibits several important features. 
First, because the physical electron is formed as a composite object consisting of a chargon and a gapped magnon with the energy $\pm\omega_{\mathbf q}$, each dispersive band splits into two branches, {as discussed in Appendix \ref{hole_pocket_analysis}}.}
{Second, a diffusive Fermi pocket located in the vicinity of $(\pi/2,\pi/2)$-point is transformed into the arc-shaped feature. Both features become even more pronounced and clearly visible as the temperature is lowered; see the discussion 
below.}

\begin{figure*}[t]
    \centering
    \includegraphics{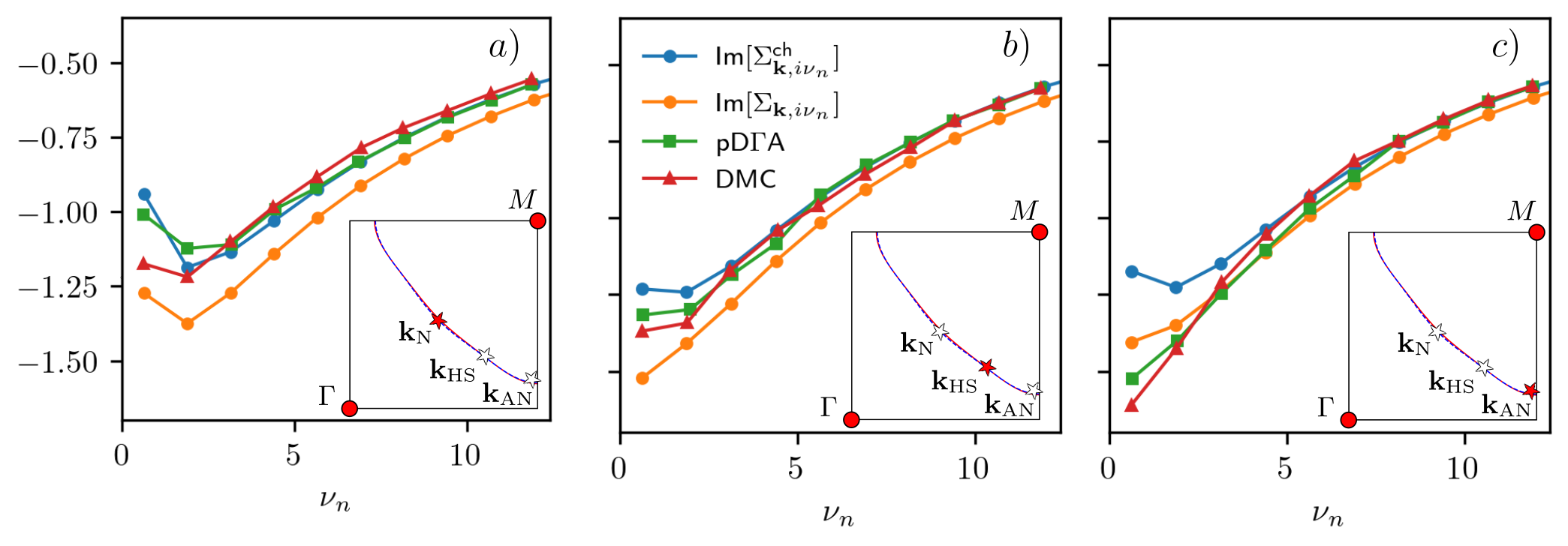}
    \caption{The imaginary parts of the chargon self-energy $\Sigma^{\mathrm{ch}}_{\mathbf{k},i\nu_n}$ and the electron self-energy $\Sigma_{\mathbf{k},i\nu_n}$, together with the diagrammatic Monte Carlo (DMC) from Ref.~\onlinecite{Wu2017} and parquet D$\Gamma$A (pD$\Gamma$A) from Ref.~\onlinecite{Kugler2025}, are shown along the imaginary-frequency axis for the following momentum points: (a) the nodal point $\mathbf{k}_{\mathrm{N}}=(1.47,1.47)$, (b) the hot-spot point $\mathbf{k}_{\mathrm{HS}}=(2.26,0.88)$, and (c) the antinodal point $\mathbf{k}_{\mathrm{AN}}=(3.04,0.49)$. The insets show the chargon ($\mu-\epsilon_{\mathbf{k}}-\text{Re}\,\Sigma^{\mathrm{ch}}_{\mathbf{k},i\nu_1}=0$, red solid line) and electron ($\mu-\epsilon_{\mathbf{k}}-\text{Re}\,\Sigma_{\mathbf{k},i\nu_1}=0$, blue dashed line) Fermi surfaces (which almost coincide) and the position of the respective points. }
    \label{sigma_block}
\end{figure*}

{If the dynamics of chargons was treated within a static mean-field approach, such an asymmetry in the damping between the inner and outer regions could not be captured. It is therefore essential to include correlation effects in order to account for the asymmetric formation of Fermi arcs.}
{As it is discussed in  Refs.~\onlinecite{OurFirst,DMFT_Incomm1} the asymmetric damping on the outer side of the hole pockets appears already in DMFT calculations at higher hole concentrations, closer to optimal doping. In contrast, at the lower doping considered in the present work, the asymmetry of the hole pockets in pure DMFT, corresponding to the chargon sector, is weak. Therefore, corrections associated with the finite correlation length of magnetic fluctuations are essential for describing the suppression of quasiparticles on the outer side of the hole pocket.}

\begin{figure}[b]
    \centering
    \includegraphics{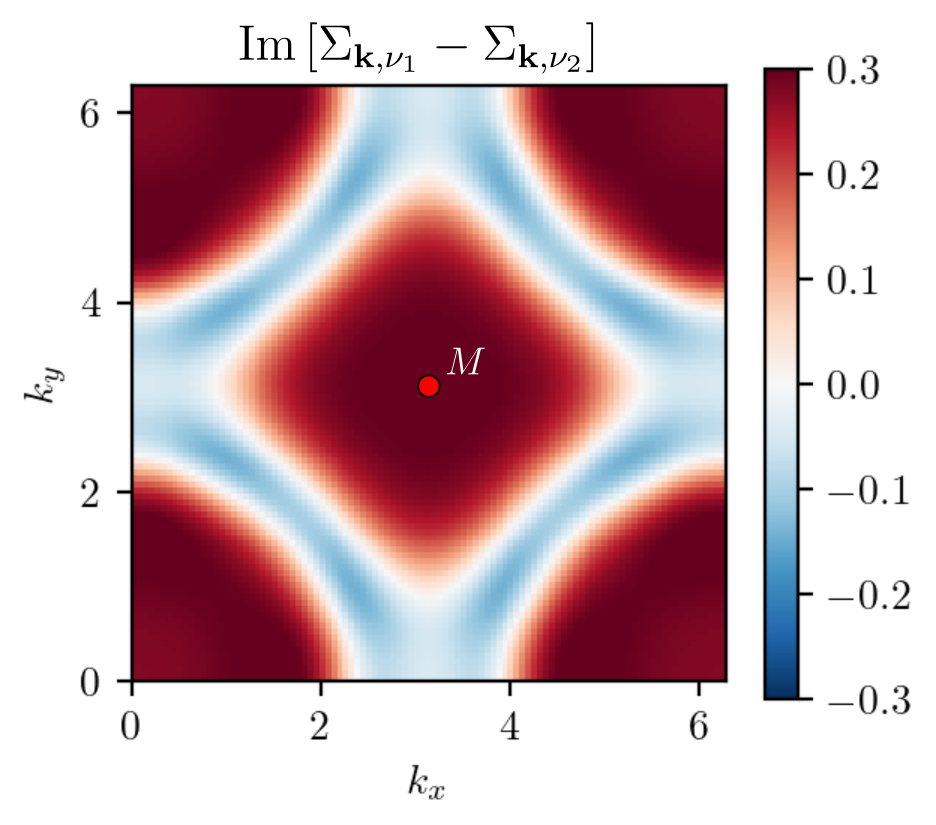}
    \caption{Imaginary part of self-energies differences 
of electrons $\Sigma_{\mathbf{k},\nu_1} - \Sigma_{\mathbf{k},\nu_2}$ as a function of wave vector $\mathbf{k}$ for $x=0.04$ for $T=0.2 t$. 
}
\label{self_energy_difference}
\end{figure}

Using the spectral density in Eq.~(\ref{A_g_convolution}), we also obtain the electron Green's function $G_{\mathbf{k},i\nu_n}$ and the electron self-energy $\Sigma_{\mathbf{k},i\nu_n}$ on the imaginary-frequency axis using the Dyson equation.
}
{Fig.~\ref{sigma_block} compares the chargon and electron self-energies as functions of imaginary frequency at $\mathbf{k}$ points located close to the surfaces where the real parts of the corresponding Green's functions change sign. One can see that the approach employed in this work, based on a fractionalized electron state, leads to a substantial enhancement of the electron damping. Comparison with the data of Refs.~\onlinecite{Wu2017,Kugler2025} shows that this enhancement is captured correctly in the antinodal direction but is somewhat overestimated in the nodal direction. 
}

\begin{figure*}[t]
    \centering
    \includegraphics[width=0.8\linewidth]{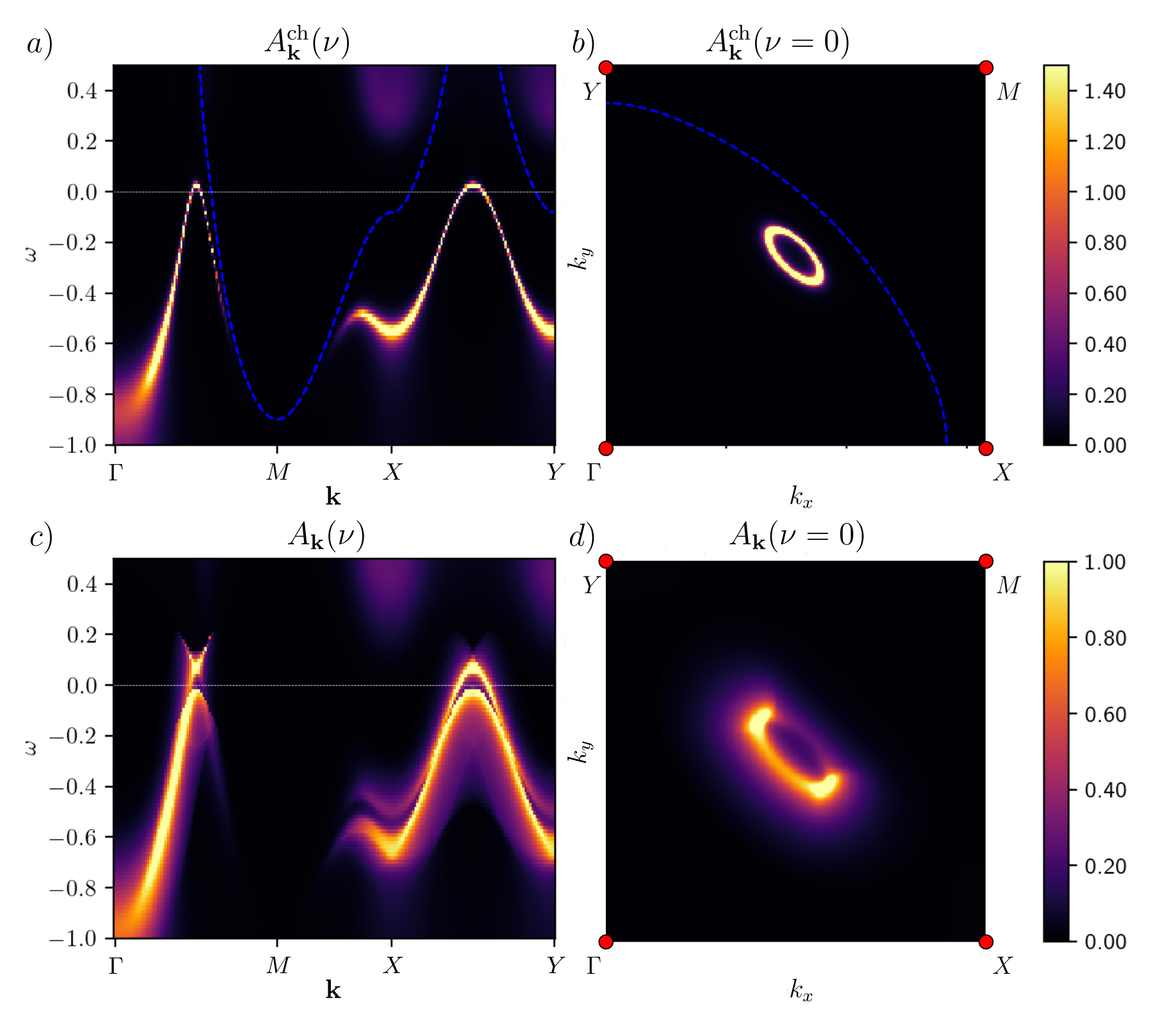}
    \caption{The same as Fig. \ref{beta_5_spectral_results} for $T=0.1t$. 
    }
    \label{beta_10_spectral_results}
\end{figure*}

The results for the difference of the imaginary parts of the self-energies ${\rm Im}(\Sigma_{\mathbf{k},\nu_1} - \Sigma_{\mathbf{k},\nu_2})$ of electrons at the first and second Matsubara frequencies $\nu_{1,2}$ as functions of $\mathbf{k}$ are shown for $T=0.2t$ in Fig.~\ref{self_energy_difference}. Similarly to the chargon self-energies, 
the non-Fermi-liquid region, where this difference is negative, is concentrated along the line defined by Eq.~(\ref{self_energy_poles_equation}).
For the physical self-energies the non-Fermi-liquid behavior becomes more pronounced, and the corresponding non-Fermi-liquid region broadens, fully extending to the $X$ and $Y$ points. Remarkably, such behavior of the self-energy on the imaginary frequency axis, as well as the corresponding location of the Luttinger surface, is in qualitative agreement with numerical results obtained within diagrammatic extensions of DMFT in the paramagnetic phase~\cite{Krien2022,Kugler2025}.

At lower temperature $T=0.1 t$ (see Fig. \ref{beta_10_spectral_results} a,b) {the chargons spectral functions become more coherent}. 
Nevertheless, these excitations are completely destroyed in the broad momentum range near $M$ point due to the singular behavior of the self-energy, see Eq.~(\ref{main_general_SE}), along the curve defined by Eq.~(\ref{self_energy_poles_equation}).
In the vicinity of $(\pi/2,\pi/2)$, a well-defined hole pocket is formed (see Fig.~\ref{beta_10_spectral_results} b). Thus, the Fermi surface of chargons is significantly deformed from its non-interacting shape by the strong momentum dependence of the self-energy~(\ref{main_general_SE}) and spectral features at higher temperatures in the nodal direction develop into the hole pockets with decreasing temperature. 

At $T=0.1 t$ we find $\rho=0.028$, $\chi=0.207$, obtaining $\omega_0 \approx 0.04\,t$ and $\xi/a \approx 9.8$.
{As in the results for $T=0.2t$ shown in Fig.~\ref{beta_5_spectral_results} c,d), we observe (see Figs. \ref{beta_10_spectral_results} c,d) a redistribution of spectral weight and band splitting, leading to the opening of a gap at the Fermi level in the region where the hole pocket was previously present.
The gap opens much more efficiently on the outer side of the hole pocket than on the inner side. On the inner side, the gap-opening effect is partially compensated by a transfer of spectral weight from deeper parts of the band below the Fermi level. This does not occur on the outer side because, closer to the $M$ point, the hole band is already destroyed and the corresponding spectral weight is absent. The imaginary part of the difference of the self-energies at the first and second Matsubara frequencies  at $T=0.1t$ looks similar to Fig. \ref{self_energy_difference}. The presence of the wide non-Fermi-liquid region in electronic self-energies explains, in a different way, the destruction of the outer side of the hole pocket since it falls onto the `non-Fermi-liquid' part of  momentum space with ${\rm Im}(\Sigma_{\mathbf{k},\nu_1} - \Sigma_{\mathbf{k},\nu_2})<0$.
We also observe that a significant amount of spectral weight remains on the parts of the hole pockets lying along the $X$--$Y$ line. As discussed in Appendix~\ref{hole_pocket_analysis}, we attribute this effect to the significantly larger effective mass of holes in this direction compared with the nodal direction, which makes these states more robust against suppression by bosonic excitations.}

\section{Conclusion}
{In the present paper we have considered the pseudogap state of the single-band Hubbard model on the square lattice within an SU(2) gauge theory, where physical electrons are represented in terms of fractionalized chargons and spinons. We have shown that the long-range magnetic order (treated within DMFT approach) yields formation of the Luttinger surface and the reconstruction of the electronic spectrum of chargons. This results in the pseudogap state of chargon subsystem, in which, at small hole doping, the single-particle excitations form hole pockets in the nodal directions. These pockets, however, do not show a pronounced asymmetry of the spectral weight between the inner and outer sides with respect to the center of the Brillouin zone.}

{At the same time, passing from chargons to physical electrons, considered with account of the spinon dynamics, leads to substantial changes in the electronic spectrum. In particular, due to the anisotropy of the effective hole mass in the pocket and the asymmetric damping of quasiparticles in the states below the Fermi level, the hole pocket is transformed into the Fermi-arc-like structure, with the spectral weight being strongly suppressed on the outer side of the pocket. This behavior is observed at different temperatures, while the spectral features become much more pronounced with decreasing temperature.}

{The obtained asymmetry of the electronic Fermi surfaces is similar to those observed experimentally in spectroscopic studies of high-$T_c$ cuprate superconductors. The considered mechanism of formation of arc-like Fermi surfaces takes place in the underdoped regime at small hole doping, where DMFT alone is not sufficient to explain the asymmetry of the spectral weight on the inner and outer sides of the hole pockets.}

{We note that the uniform external magnetic field is expected to affect predominantly the chargon dynamics, while leaving the spinon sector essentially unchanged. Therefore, we expect that the oscillations induced by such a field are determined by the chargon hole pockets. At the same time, the presence of spinons is not expected to substantially modify these quantum oscillations of the electronic spectral density; they can rather reduce the oscillation amplitude by smearing these features. In contrast, as we demonstrate in the present work, spinons directly affect photoemission measurements of the electronic spectrum, changing the observed spectral features of physical electrons. Therefore, the obtained results can be considered as a good starting point for further description of high-$T_c$ superconductors.}

\label{conclustion_sec}

\newpage
\section*{Acknowledgements}
The authors are grateful to A. I. Lichtenstein, E. A. Stepanov, and F. Kugler for discussions. The calculations of spectral functions are supported by the Russian
Science Foundation (Grant No. 24-12-00186). Development of DMFT and nonlocal extension programs is supported in part by the state assignments of the Ministry of
Science and Higher Education of the Russian Federation
for the IMP UB RAS and FSMG-2025-0005 for MIPT. I. A. G.
also acknowledges financial support of the derivation of the mean-field equations for chargon and spinon subsystem from BASIS Foundation
(Grant No. 24-1-5-152-1). 

\begin{widetext}
\appendix
\renewcommand\thefigure{A\arabic{figure}}
\setcounter{figure}{0}

\section{Mean-field study of {the spinon subsystem}}
\label{z_field_mean_field_derivation}

We start from the action in the form (\ref{A_action}), where we do not assume spin stiffness $\mathcal{J}^{ab}_{\mu,x; \nu,x'}$ to be local. Here we consider antiferromagnetic case where we can choose is to have a general form
\begin{equation}
    \mathcal{J}^{ab}_{\mu,x; \nu,x'} =  
\begin{pmatrix}  
J_{\mu,x; \nu,x'} & 0 & 0 \\  
0 & J_{\mu,x; \nu,x'} & 0 \\  
0 & 0 & 0  
\end{pmatrix}
\end{equation}

In the case of nonlocal stiffness, the standard derivation of the mapping to the nonlinear $\sigma$-model and its $\mathrm{CP}^1$ formulation is not straightforward, since the relation $\mathcal{R}_x^\dagger \mathcal{R}_x = \mathbf{1}$ does not generalize to the nonlocal case with $x \neq x'$. In the nonlocal case, it is therefore convenient to rewrite the action~(\ref{A_action}) directly in terms of the bosonic field $z_x$, $z_x^\dagger$ using the parametrization (\ref{R_parametrization}). Then the Eq.~(\ref{A_direct_z}) holds, which can be directly substituted into the action~(\ref{A_action}). Note that the parametrization (\ref{R_parametrization}) differs from the more common $\mathrm{CP}^1$ parametrization \cite{}, where $n_x^a = z_x^\dagger \sigma^a z_x$, $a=x,y,z$.

After the substitution and usage of identity
    $\sigma^{x}_{\sigma\sigma'} \sigma^{x}_{\sigma''\sigma'''} +
\sigma^{y}_{\sigma\sigma'} \sigma^{y}_{\sigma''\sigma'''} =
2 \delta_{\sigma,-\sigma''} \delta_{\sigma,-\sigma'} \delta_{\sigma'',-\sigma'''},$
we obtain the action in the form
\begin{align}
    S^{\text{eff}}[z] = \sum_{x,x'} \sum_{\mu,\nu} J_{\mu,x; \nu,x'} \sum_{\sigma}
    \Big[
    (\partial_{\mu,x} z_{\sigma,x}^*) z_{-\sigma,x} z_{-\sigma,x'}^* (\partial_{\nu,x'} z_{\sigma,x'}) +
z_{\sigma,x}^* (\partial_{\mu,x} z_{-\sigma,x}) (\partial_{\nu,x'} z_{-\sigma,x'}^*) z_{\sigma,x'} \notag \\ -
(\partial_{\mu,x} z_{\sigma,x}^*) z_{-\sigma,x} (\partial_{\nu,x'} z_{-\sigma,x'}^*) z_{\sigma,x'}
- z_{\sigma,x}^* (\partial_{\mu,x} z_{-\sigma,x}) z_{-\sigma,x'}^* (\partial_{\nu,x'} z_{\sigma,x'})
    \Big],
\end{align}
were the bosonic field $z_x,z_x^+$ satisfies the constraint (\ref{z_constraint}).

Next we write a mean-field approximation for this action. We apply Wick's theorem to the products of
the first-order expressions of $z_x,z_x^+$ and their derivatives and we omit all contributions to the action independent from $z_x$-field. We consider only a symmetric state without condensation of bosons, choosing the following ansatz for the mean-field propogator of the $z_z$-field
\begin{align}
    & \langle z_{\sigma,x}^* z_{\sigma',x'}\rangle=\delta_{\sigma\sigma'}\,D_{x,x'}, \\
    & \langle z_{\sigma,x}^* z_{-\sigma,x'}\rangle=
    \langle z_{\sigma,x} z_{\sigma',x'}\rangle=
     \langle z_{\sigma,x}^* z_{\sigma',x'}^*\rangle=0.
    \label{MF_propogators_rule}
\end{align}
The mean-field approximation for the action is
    $S^{\text{eff}}[z] \approx S^{(1)}[z] + S^{(2)}[z]$,
\begin{align}
    S^{(1)}[z] =  \sum_{x,x'} \sum_{\mu,\nu} J_{\mu,x; \nu,x'} \sum_{\sigma}
    &\Big[
(\partial_{\mu,x} \partial_{\nu,x'}D_{x,x'}) z_{\sigma,x} z_{\sigma,x'}^*
+ D_{x',x} (\partial_{\mu,x} z_{\sigma,x}^*) (\partial_{\nu,x'} z_{\sigma,x'}) \notag\\
&+ D_{x,x'} (\partial_{\mu,x} z_{\sigma,x}) (\partial_{\nu,x'} z_{\sigma,x'}^*) +
(\partial_{\mu,x}\partial_{\nu,x'} D_{x',x}) z_{\sigma,x}^* z_{\sigma,x'}
    \Big],\\
    S^{(2)}[z] =  \sum_{x,x'} \sum_{\mu,\nu} J_{\mu,x; \nu,x'} \sum_{\sigma}
    &\Big[
(\partial_{\mu,x} D_{x,x'}) z_{\sigma,x} (\partial_{\nu,x'} z_{\sigma,x'}^*) +
(\partial_{\nu,x'} D_{x',x}) (\partial_{\mu,x} z_{\sigma,x}^*) z_{\sigma,x'} \notag\\ &+
(\partial_{\nu,x'} D_{x,x'})  (\partial_{\mu,x} z_{\sigma,x}) z_{\sigma,x'}^* +
(\partial_{\mu,x} D_{x',x}) z_{\sigma,x}^* (\partial_{\nu,x'} z_{\sigma,x'})
    \Big].
\end{align}

Local constraint~(\ref{z_constraint}) implies that spatial variations of $z_x$ are not arbitrary, but must satisfy the following first-order differential equation, written for fixed $\sigma$:
\begin{equation}
    (\partial_{\mu,x} z_{\sigma,x}^*) z_{\sigma,x} + z_{\sigma,x}^* (\partial_{\mu,x}z_{\sigma,x}) + (\partial_{\mu,x} z_{-\sigma,x}^*) z_{-\sigma,x} + z_{-\sigma,x}^* (\partial_{\mu,x}z_{-\sigma,x}) = 0.
    \label{differential_constaint}
\end{equation}
A useful set of equations can be obtained by multiplying this equation by terms of second order in the field $z_x$, such as $(\partial_{\nu,x'} z_{-\sigma,x'}^*) z_{-\sigma,x'}$ or $(\partial_{\nu,x'} z_{-\sigma,x'}) z_{-\sigma,x'}^*$, and then reducing the resulting fourth-order expressions to second order by applying the same mean-field decoupling procedures
with the averages (\ref{MF_propogators_rule}), as used for the action itself. This procedure, together with a repeated use of the constraint~(\ref{differential_constaint}), leads to the equations
\begin{align}
    D_{x,x'} (\partial_{\mu,x}z_{-\sigma,x}) (\partial_{\nu,x'} z_{-\sigma,x'}^*) &+
(\partial_{\mu,x} \partial_{\nu,x'} D_{x',x}) z_{-\sigma,x}^* z_{-\sigma,x'} \notag\\ &+
(\partial_{\mu,x} D_{x,x'}) z_{-\sigma,x} (\partial_{\nu,x'} z_{-\sigma,x'}^*) +
(\partial_{\nu,x'} D_{x',x}) (\partial_{\mu,x} z_{-\sigma,x}^*) z_{-\sigma,x'} = 0,
\label{mf_relation_1}\\
(\partial_{\mu,x} \partial_{\nu,x'} D_{x,x'}) z_{-\sigma,x} z_{-\sigma,x'}^* &+ D_{x',x} (\partial_{\mu,x} z_{-\sigma,x}^*) (\partial_{\nu,x'} z_{-\sigma,x'}) \notag \\ &+
(\partial_{\nu,x'} D_{x,x'}) (\partial_{\mu,x}z_{-\sigma,x}) z_{-\sigma,x'}^* +
(\partial_{\mu,x} D_{x',x}) z_{-\sigma,x}^* (\partial_{\nu,x'} z_{-\sigma,x'}) = 0.
\label{mf_relation_2}
\end{align}
From these equations it follows that $S^{(1)}[z] \approx S^{(2)}[z]$ under the same assumptions under which the mean-field approach is applicable to the action (\ref{A_action}) itself. 

After passing to momentum space according to
$z_x = \sum_q e^{-i q\cdot x}\,z_q$,
the effective action takes the form
\begin{equation}
S^{\text{eff}} [z,z^+]
\approx
\sum_{\sigma} \sum_{k,q} \sum_{\mu,\nu} 2\,J_{\mu,\nu;q} z^*_{\sigma,k} 
\sum_{\alpha=\pm}
\Big[
k_{\mu} k_{\nu}+
(k_{\mu} + \alpha q_{\mu}) (k_{\nu} + \alpha q_{\nu})
\Big] D_{k+\alpha q} \, z_{\sigma,k}.
\end{equation}
Using the identity
    $k_{\mu} k_{\nu}+
(k_{\mu} + \alpha q_{\mu}) (k_{\nu} + \alpha q_{\nu}) = 
 [
(2 k_{\mu} + \alpha q_{\mu}) (2k_{\nu} + \alpha q_{\nu}) + q_{\mu} q_{\nu}
]/2$,
which applies to the terms originating from the transformed expression for~$S^{(1)}[z]$, action can be rewritten, up to corrections to the chemical potential, as
\begin{equation}
    S^{\text{eff}} [z,z^+]
\approx \sum_{k,\sigma} z^*_{k,\sigma} \Sigma_k z_{k,\sigma},
\end{equation}
where $\Sigma_k$ is defined by Eq.~(\ref{sigma_k_expression}).

Propagator and the sum-rule equation in the forms (\ref{z_propogator}) and (\ref{sum_rule_equation}) can be obtained by accounting for the constraint using an additional term in the action
    ${i} \sum_x \lambda_x (z^+_x z_x - 1)$
and a saddle point approximation 
    ${i} \lambda_x = 2 \Delta$.

Note that the relations~(\ref{mf_relation_1}) and~(\ref{mf_relation_2}) must be applied before taking the local limit, $J_{\mu,\nu;q} \propto \delta_{q,0}$, in order to recover the correct low-energy theory, which matches the known results of the NLSM representation in the local case. 

\section{Temporal spin stiffness}
\label{temporal_spin_stiffness_ap}
To assess the validity of the static temporal-stiffness approximation, we computed its full dynamical frequency dependence according to the expression~$\chi_{\omega_n} = {m^2}/({\chi^{yy}_{\omega_n,\mathbf{q} = \mathbf{Q}} \omega_n^2})$. The corresponding results are presented in Fig.~\ref{chi_wn}.
One can see that, as the temperature is lowered from $T=0.2t$ to $T=0.1t$, the stiffness increases rather sharply, while upon further cooling it remains almost unchanged. Comparison with the mean-field results of Ref.~\onlinecite{OurThird} shows that in the present DMFT calculation the frequency dependence of the spin stiffness is much weaker, whereas its high-frequency asymptotic value is substantially larger. Apparently, this reflects better applicability of the non-linear sigma model with the 
static spin-stiffness approximation used in the main part of the paper, than for the mean-field theory approach.

\begin{figure*}[h]
    \centering
\includegraphics{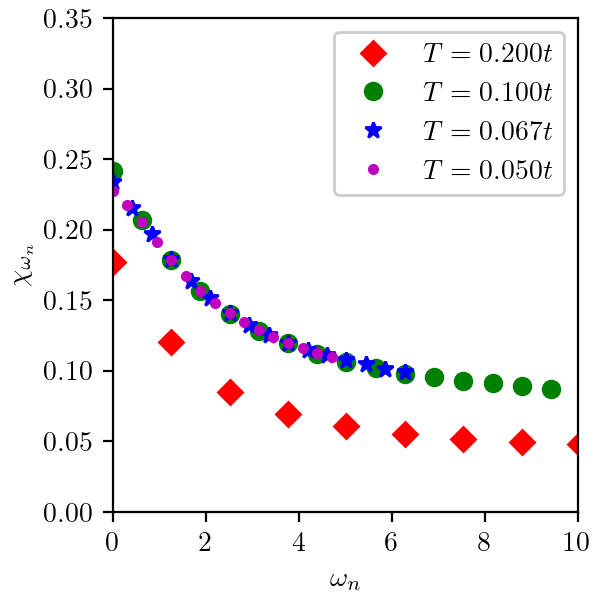}
    \caption{Dynamic temporal spin stiffness $\chi_{\omega_n}$ as functions of bosonic imaginary frequency $\omega_n$ for $U=5.6t$ at various temperatures $T$.}
\label{chi_wn}
\end{figure*}

\section{Influence of magnetic fluctuations on the hole pocket}
\label{hole_pocket_analysis}
We can qualitatively understand the effect of short-ranged magnetic fluctuations on the redistribution of spectral weight in the vicinity of the Fermi level by considering a simple model spectrum. As such a model, we consider an anisotropic hole pocket with spectral density and dispersion given by $A^{\text{ch}}_{\mathbf{k}}(\nu) = \delta(\nu - \xi^{\text{ch}}_{\mathbf{k}})$ with
\begin{align}
    & \xi^{\text{ch}}_{\mathbf{k}} = \mu - \frac{|(\mathbf{k} - \mathbf{k}_0,\mathbf{e_{\parallel}})|^2}{2 m^*_{\parallel}} - \frac{|(\mathbf{k} - \mathbf{k}_0,\mathbf{e_{\perp}})|^2}{2m^*_{\perp}},
    \label{hole_pocket_model_dispresion_2}
\end{align}
where $\mathbf{e}_{\parallel}$ and $\mathbf{e}_{\perp}$ are unit vectors in the directions $\Gamma\to M$ and $X \to Y$ respectively, $\mathbf{k}_0$ is a center of the hole pocket.

According to the Eq. (\ref{A_g_convolution}), spinon--chargon pairs form two excitation branches, with the upper branch corresponding to the energy $\nu=\xi_{\mathbf{k}-\mathbf{q}}+\omega_{\mathbf q}$. Expansion in momentum $q$ yields 
\begin{equation}
\nu =
-{\mathbf v}_{F,{\mathbf k}} {\mathbf q}
+\xi_{\mathbf{k}}+\omega_{q}-\frac{1}{2}\left(\frac{({\mathbf e}_{\parallel} {\mathbf q})^2}{m^*_\parallel}+\frac{({\mathbf e}_{\perp} {\mathbf q})^2}{m^*_\perp}\right),
\label{nu_curve_family}
\end{equation}
where ${\mathbf v}_{F,{\mathbf k}}=(\mathbf{k}- \mathbf{k}_0,\mathbf{e_{\parallel}})\mathbf{e_{\parallel}}/{m^*_{\parallel}}+(\mathbf{k}- \mathbf{k}_0,\mathbf{e_{\perp}})\mathbf{e_{\perp}}/{m^*_{\perp}}$ is the Fermi velocity. At finite magnetic gap $\Delta$, spinons, similarly to chargons, can be assigned an effective mass $m_s = {\omega_0}/{v^2}$. This yields
\begin{align}
\nu&=\frac{1}{2}\sum_{i=\parallel,\perp}
\left[\left(\frac{1}{m_s}-\frac{1}{m^*_i}\right)\left(({\mathbf e}_{i} {\mathbf q})-\frac{m_s (\mathbf{k}- \mathbf{k}_0,\mathbf{e_{i}})}{m^*_i-m_s}\right)^2
-\frac{m_s}{m^*_{i}}\frac{(\mathbf{k}- \mathbf{k}_0,\mathbf{e_{i}})^2}{m^*_i-m_s}
\right]
+\xi_{\mathbf{k}}+\omega_{0}.
\label{nu_curve_family}
\end{align}
After averaging over all relative momenta $\mathbf q$ of the spinon--chargon pairs, the first term in the square brackets leads to a broadening of the spectral weight, whereas the dominant $k$-dependence of the upper excitation branch is determined by the second term,which yields the effective dispersion
\begin{equation}
\tilde \xi_{\mathbf{k}} = \mu + \omega_0- 
\frac{|(\mathbf{k} - \mathbf{k}_0,\mathbf{e_{\parallel}})|^2}{2(m^*_{\parallel} - m_s)} - 
\frac{|(\mathbf{k} - \mathbf{k}_0,\mathbf{e_{\perp}})|^2}{2(m^*_{\perp} - m_s)}.
\end{equation}
Thus the effective mass of the excitations in the upper branch is determined by the quantity $m^* - m_s$.
If the effective mass of holes is smaller than the effective mass of spinons, the upper excitation branch bends toward higher energies as $\kappa = |\mathbf{k} - \mathbf{k}_0|$ increases, and the gap does not close; in this case, the hole pocket evolves into a true gap. By contrast, if the effective mass exceeds the $m_s$, the upper branch bends downward with increasing $\kappa$ and, at some finite $\kappa$, crosses the Fermi level again. In this case, the hole pocket expands without transforming into a full gap.

The resulting spectral-weight distribution is sensitive to the anisotropy of the hole pocket and of the effective mass in different directions. This consideration was checked directly by computing the electronic spectral function~(\ref{A_g_convolution}) for the hole-pocket model~\ref{hole_pocket_model_dispresion_2}). The parameters were estimated from the spectrum shown in Fig.~\ref{beta_10_spectral_results}(a,b): $k_{0,x} = k_{0,y} \approx \pi / 2$, $\mu \approx 0.026$, $m^*_{\parallel} \approx 0.2935$ in the $\Gamma\to M$ direction, and $m^*_{\perp} \approx 1.5028$ in the $X \to Y$ direction. We used an artificially broadened chargon spectral function with broadening $\eta=0.001$, as well as the spin stiffness and magnetic gap $\Delta$ corresponding to the same state at $x=4\%$ and $T=0.1t$. The results are shown in Fig.~\ref{hole_pocket_spectral_results}. One can see that the anisotropy of the effective masses indeed leads to different behavior of the resulting spectral weight along the $\Gamma\to M$ and $X \to Y$ directions.

The same qualitative difference is observed in the full numerical results presented in Figs.~\ref{beta_5_spectral_results}(c,d) and~\ref{beta_10_spectral_results}(c,d). In particular, the behavior of the spectral line along the $\Gamma \to M$ and $X \to Y$ directions is qualitatively different. Along the $X \to Y$ direction, no full gap is formed; instead, the hole pocket simply expands. This difference can therefore be attributed to the different effective hole masses along these two directions relative to the gapped magnetic excitations.
At the same time, the results shown in Fig.~\ref{beta_10_spectral_results} demonstrate that obtaining an arc-like Fermi surface requires taking into account correlation effects absent in the hole-pocket model. These effects strongly affect the amount of spectral weight transferred from below the Fermi level by spinons.

\begin{figure*}[t]
    \centering
    \includegraphics{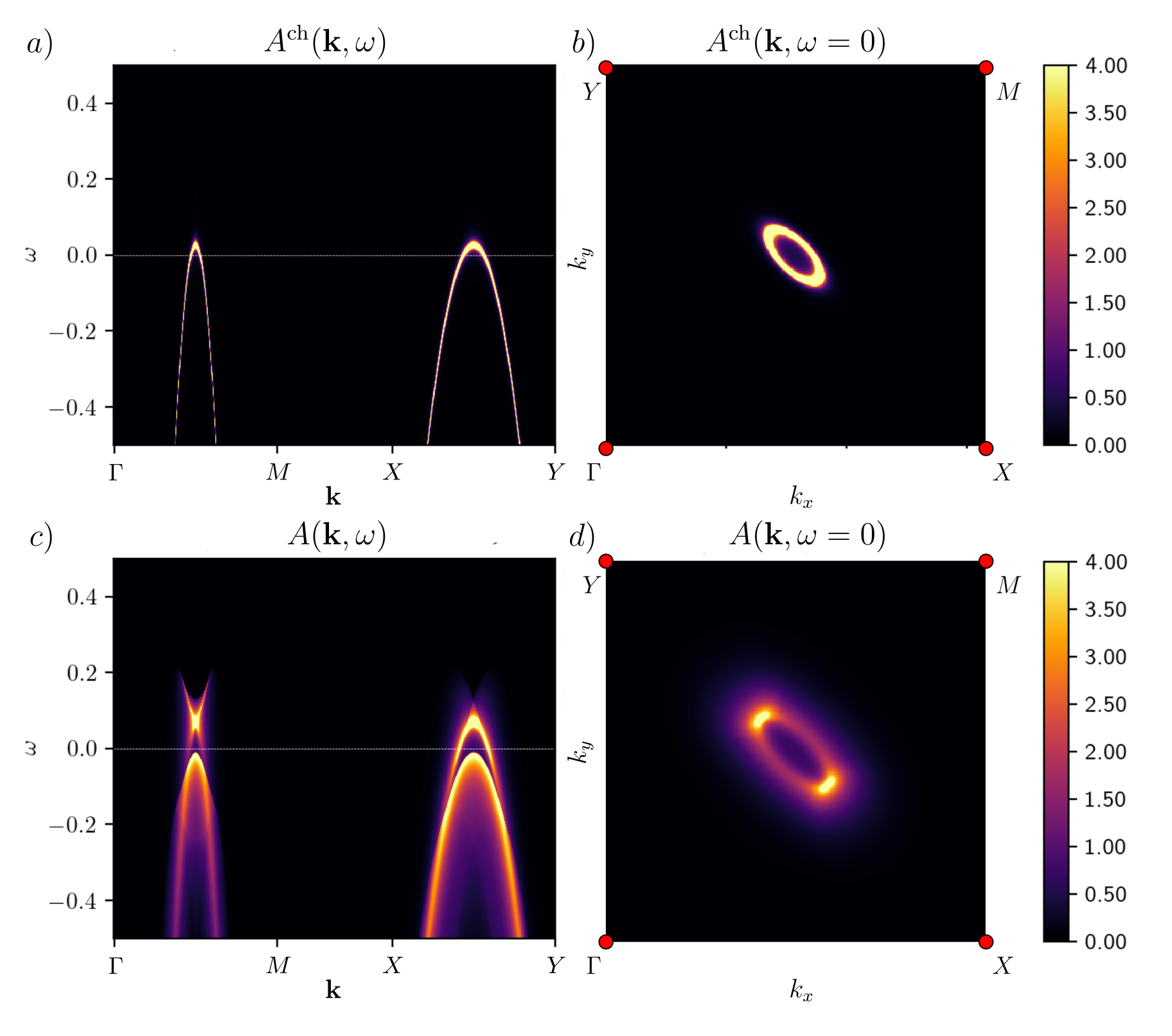}
    \caption{\centering  a-b) Spectral density $A^{\text{ch}}_{\mathbf{k}}(\omega)$ and Fermi surface $A^{\text{ch}}_\mathbf{k}(\omega = 0)$, c-d) same quantities $A(\mathbf{k},\omega)$ and $A_\mathbf{k}(\omega = 0)$ for hole pocket model of chargons in a state $x=0.040$ at $T =0.1 t$ Hole's dispersion was artificially broadened with broadening $\eta=0.001$.
    }
    \label{hole_pocket_spectral_results}
\end{figure*}

\end{widetext}


\begin{thebibliography}{99}



\bibitem{Berthier1996Review} C. Berthier, M. H. Julien, M. Horvatić, Y. Berthier, NMR studies of the normal state of high temperature superconductors, \href{https://dx.doi.org/10.1051/jp1:1996209}{J. Phys. I France \textbf{6}, 2205 (1996)}.

\bibitem{Asayama1996Review} K. Asayama, Y. Kitaoka, G. Q. Zheng, K. Ishida, NMR studies of high $T_c$ superconductors, \href{https://dx.doi.org/10.1016/0079-6565(95)01025-4}{Prog. Nucl. Magn. Reson. Spectrosc. \textbf{28}, 221 (1996)}.

\bibitem{Plakida} N. Plakida, \textit{High-Temperature Cuprate Superconductors} (Springer, Heidelberg, 2010).


\bibitem{Ando2004} Y. Ando, Y. Kurita, S. Komiya, S. Ono, K. Segawa, Evolution of the Hall coefficient and the peculiar electronic structure of the cuprate superconductors, \href{https://dx.doi.org/10.1103/PhysRevLett.92.197001}{Phys. Rev. Lett. \textbf{92}, 197001 (2004)}.

\bibitem{Ando2004PhaseDiagram} Y. Ando, S. Komiya, K. Segawa, S. Ono, Y. Kurita, Electronic phase diagram of high-$T_c$ cuprate superconductors from a mapping of the in-plane resistivity curvature, \href{https://dx.doi.org/10.1103/PhysRevLett.93.267001}{Phys. Rev. Lett. \textbf{93}, 267001 (2004)}.

\bibitem{Ono2007} S. Ono, S. Komiya, Y. Ando, Strong charge fluctuations manifested in the high-temperature Hall coefficient of high-$T_c$ cuprates, \href{https://dx.doi.org/10.1103/PhysRevB.75.024515}{Phys. Rev. B \textbf{75}, 024515 (2007)}.


\bibitem{ARPESReview_Damascelli2003} A. Damascelli, Z.-X. Shen, and Z. Hussain, Angle-resolved photoemission studies of the cuprate superconductors, \href{https://doi.org/10.1103/RevModPhys.75.473}{Rev. Mod. Phys. \textbf{75}, 473 (2003)}.

\bibitem{ARPES_Bi2201_DopingEvolution_Hashimoto2008} M. Hashimoto, T. Yoshida, H. Yagi, M. Takizawa, A. Fujimori, M. Kubota, K. Ono, K. Tanaka, D. H. Lu, Z.-X. Shen, S. Ono, and Y. Ando, Doping evolution of the electronic structure in the single-layer cuprates Bi$_2$Sr$_{2-x}$La$_x$CuO$_{6+\delta}$: Comparison with other single-layer cuprates, \href{https://doi.org/10.1103/PhysRevB.77.094516}{Phys. Rev. B \textbf{77}, 094516 (2008)}.


\bibitem{ARPES_LSCO_UnderlyingFS_Yoshida2006} T. Yoshida, X. J. Zhou, K. Tanaka, W. L. Yang, Z. Hussain, Z.-X. Shen, A. Fujimori, S. Komiya, Y. Ando, H. Eisaki, T. Kakeshita, and S. Uchida, Systematic doping evolution of the underlying Fermi surface of La$_{2-x}$Sr$_x$CuO$_4$, \href{https://doi.org/10.1103/PhysRevB.74.224510}{Phys. Rev. B \textbf{74}, 224510 (2006)}.

\bibitem{ARPES_NaCCOC_Shen2005} K. M. Shen, F. Ronning, D. H. Lu, F. Baumberger, N. J. C. Ingle, W. S. Lee, W. Meevasana, Y. Kohsaka, M. Azuma, M. Takano, H. Takagi, and Z.-X. Shen, Nodal quasiparticles and antinodal charge ordering in Ca$_{2-x}$Na$_x$CuO$_2$Cl$_2$, \href{https://doi.org/10.1126/science.1103627}{Science \textbf{307}, 901 (2005)}.

\bibitem{Norman1998} M. R. Norman, H. Ding, M. Randeria, J. C. Campuzano, T. Yokoya, T. Takeuchi, T. Takahashi, T. Mochiku, K. Kadowaki, P. Guptasarma, D. G. Hinks, Destruction of the Fermi surface in underdoped high-$T_c$ superconductors, \href{https://dx.doi.org/10.1038/32366}{Nature \textbf{392}, 157 (1998)}.

\bibitem{Kanigel2006} A. Kanigel, M. R. Norman, M. Randeria, U. Chatterjee, S. Souma, A. Kaminski, H. M. Fretwell, S. Rosenkranz, M. Shi, T. Sato, et. al., Evolution of the pseudogap from Fermi arcs to the nodal liquid, \href{https://dx.doi.org/10.1038/nphys334}{Nature Physics \textbf{2}, 447 (2006)}.

\bibitem{DoironLeyraud2007} N. Doiron-Leyraud, C. Proust, D. LeBoeuf, J. Levallois, J.-B. Bonnemaison, R. Liang, D. A. Bonn, W. N. Hardy, L. Taillefer, Quantum oscillations and the Fermi surface in an underdoped high-$T_c$ superconductor, \href{https://dx.doi.org/10.1038/nature05872}{Nature \textbf{447}, 565 (2007)}.

\bibitem{LeBoeuf2007} D. LeBoeuf, N. Doiron-Leyraud, J. Levallois, R. Daou, J.-B. Bonnemaison, N. E. Hussey, L. Balicas, B. J. Ramshaw, R. Liang, D. A. Bonn, W. N. Hardy, S. Adachi, C. Proust, L. Taillefer, Electron pockets in the Fermi surface of hole-doped high-$T_c$ superconductors, \href{https://dx.doi.org/10.1038/nature06332}{Nature \textbf{450}, 533 (2007)}.

\bibitem{DoironLeyraud2015} N. Doiron-Leyraud, S. Badoux, S. René de Cotret, D. LeBoeuf, N. E. Hussey, H. Chang, B. J. Ramshaw, R. Liang, D. A. Bonn, W. N. Hardy, L. Taillefer, Evidence for a small hole pocket in the Fermi surface of underdoped YBa$_2$Cu$_3$O$_y$, \href{https://dx.doi.org/10.1038/ncomms7034}{Nature Communications \textbf{6}, 6034 (2015)}.

\bibitem{Yelland2008} E. A. Yelland, J. Singleton, C. H. Mielke, N. Harrison, F. F. Balakirev, B. Dabrowski, J. R. Cooper, Quantum oscillations in the underdoped cuprate YBa$_2$Cu$_4$O$_8$, \href{https://dx.doi.org/10.1103/PhysRevLett.100.047003}{Phys. Rev. Lett. \textbf{100}, 047003 (2008)}.

\bibitem{Sebastian2011} S. E. Sebastian, N. Harrison, M. M. Altarawneh, R. Liang, D. A. Bonn, W. N. Hardy, G. G. Lonzarich, Chemical potential oscillations from nodal Fermi surface pocket in the underdoped high-temperature superconductor YBa$_2$Cu$_3$O$_{6+x}$, \href{https://dx.doi.org/10.1038/ncomms1468}{Nature Communications \textbf{2}, 471 (2011)}.

\bibitem{Barisic2013} N. Barišić, S. Badoux, M. K. Chan, C. Dorow, W. Tabis, B. Vignolle, G. Yu, J. Béard, X. Zhao, C. Proust, and M. Greven, Universal quantum oscillations in the underdoped cuprate superconductors, \href{https://dx.doi.org/10.1038/nphys2792}{Nature Physics \textbf{9}, 761 (2013)}.

\bibitem{Chan2016} M. K. Chan, N. Harrison, R. D. McDonald, B. J. Ramshaw, K. A. Modic, N. Barišić, M. Greven, Single reconstructed Fermi surface pocket in an underdoped single-layer cuprate superconductor, \href{https://dx.doi.org/10.1038/ncomms12244}{Nature Communications \textbf{7}, 12244 (2016)}.

\bibitem{SebastianProust2015} S. E. Sebastian and C. Proust, Quantum oscillations in hole-doped cuprates, \href{https://dx.doi.org/10.1146/annurev-conmatphys-030212-184305}{Annual Review of Condensed Matter Physics \textbf{6}, 411 (2015)}.

\bibitem{ARPESPockets} H.-B. Yang, J. D. Rameau, Z.-H. Pan, G. D. Gu, P. D. Johnson, H. Claus, D. G. Hinks, and T. E. Kidd, Reconstructed Fermi Surface of Underdoped Bi$_2$Sr$_2$CaCu$_2$O$_{8+\delta}$ Cuprate Superconductors, \href{http://dx.doi.org/10.1103/PhysRevLett.107.047003}{Phys. Rev. Lett. {\bf 107}, 047003 (2011)}.

\bibitem{ARPES_FermiArcsPockets_Meng2009} J. Meng, G. Liu, W. Zhang, L. Zhao, H. Liu, X. Jia, D. Mu, S. Liu, X. Dong, W. Lu, G. Wang, Y. Zhou, Y. Zhu, X. Wang, Z. Xu, C. Chen, and X. J. Zhou, Coexistence of Fermi arcs and Fermi pockets in a high-$T_c$ copper oxide superconductor, \href{https://doi.org/10.1038/nature08521}{Nature \textbf{462}, 335 (2009)}.

\bibitem{Sebastian2008} S. E. Sebastian, N. Harrison, E. Palm, T. P. Murphy, C. H. Mielke, R. Liang, D. A. Bonn, W. N. Hardy, G. G. Lonzarich, A multi-component Fermi surface in the vortex state of an underdoped high-$T_c$ superconductor, \href{https://doi.org/10.1038/nature07095}{Nature \textbf{454}, 200 (2008)}

\bibitem{Millis2007} A. J. Millis, M. R. Norman, Antiphase stripe order as the origin of electron pockets observed in 1/8-hole-doped cuprates, \href{https://doi.org/10.1103/PhysRevB.76.220503}{Phys. Rev. B \textbf{76}, 220503(R) (2007)}

\bibitem{Chakravarty2008} S. Chakravarty, H.-Y. Kee, Fermi pockets and quantum oscillations of the Hall coefficient in high-temperature superconductors, \href{https://doi.org/10.1073/pnas.0804002105}{PNAS \textbf{105}, 8835 (2008)}.


\bibitem{Millis1990} A.~J.~Millis, H.~Monien, and D.~Pines, Phenomenological model of nuclear relaxation in the normal state of YBa$_2$Cu$_3$O$_7$, \href{https://doi.org/10.1103/PhysRevB.42.167}{{Phys.\ Rev.\ B} \textbf{42}, 167 (1990).}

\bibitem{Millis1994} A.~J.~Millis, Spin fluctuations in high-temperature superconductors, \href{https://doi.org/10.1103/PhysRevB.50.16052}{{Phys.\ Rev.\ B} \textbf{50}, 16052--16055 (1994).}

\bibitem{Zha1996} Y.~Zha, V.~Barzykin, and D.~Pines, NMR and neutron-scattering experiments on the cuprate superconductors: A critical reexamination, \href{https://doi.org/10.1103/PhysRevB.54.7561}{{Phys.\ Rev.\ B} \textbf{54}, 7561--7574 (1996).}

\bibitem{Chubukov1996} A.~V.~Chubukov, D.~Pines, and B.~P.~Stojkovi\'c, Temperature crossovers in cuprates, \href{https://doi.org/10.1088/0953-8984/8/48/021}{{J.\ Phys.: Condens.\ Matter} \textbf{8}, 10017--10036 (1996).}

\bibitem{Schmalian1999} J.~Schmalian, D.~Pines, and B.~Stojkovi\'c, Microscopic theory of weak pseudogap behavior in the underdoped cuprate superconductors: General theory and quasiparticle properties, \href{https://doi.org/10.1103/PhysRevB.60.667}{{Phys.\ Rev.\ B} \textbf{60}, 667 (1999).}

\bibitem{Kuchinskii1999} E.~Z.~Kuchinskii and M.~V.~Sadovskii, Models of the pseudogap state of two-dimensional systems, \href{https://doi.org/10.1134/1.558879}{{JETP} \textbf{88}, 968 (1999).}

\bibitem{Onufrieva1999} F. Onufrieva, P. Pfeuty, and M. Kiselev, New Scenario for High-Tc Cuprates: Electronic Topological Transition as a Motor for Anomalies in the Underdoped Regime, \href{https://doi.org/10.1103/PhysRevLett.82.2370}{{Phys.\ Rev.\ Lett.} \textbf{82}, 2370 (1999).}


\bibitem{SokolPines1993} A.~Sokol and D.~Pines, Toward a unified magnetic phase diagram of the cuprate superconductors, \href{https://doi.org/10.1103/PhysRevLett.71.2813}{{Phys.\ Rev.\ Lett.} \textbf{71}, 2813 (1993).}

\bibitem{ChubukovPinesStojkovic1995} A.~V.~Chubukov, D.~Pines, and B.~P.~Stojkovi\'c, Crossover and scaling in a nearly antiferromagnetic Fermi liquid in two dimensions, \href{https://doi.org/10.1103/PhysRevB.51.14874}{{Phys.\ Rev.\ B} \textbf{51}, 14874 (1995).}

\bibitem{Wu2017} W.~Wu, M.~Ferrero, A.~Georges, and E.~Kozik, Controlling Feynman diagrammatic expansions: Physical nature of the pseudogap in the two-dimensional Hubbard model, \href{https://doi.org/10.1103/PhysRevB.96.041105}{{Phys.\ Rev.\ B} \textbf{96}, 041105(R) (2017).}

\bibitem{Sachdev2} M. S. Scheurer, S. Chatterjee, W. Wu, M. Ferrero, A. Georges, and S. Sachdev, Topological order in the pseudogap metal, \href{https://doi.org/10.1073/pnas.1720580115}{PNAS \textbf{115}, E3665 (2018)}. 

\bibitem{Wu2018} W. Wu, M. S. Scheurer, S. Chatterjee, S. Sachdev, A. Georges, and M. Ferrero, Pseudogap and Fermi-Surface Topology in the Two-Dimensional Hubbard Model, \href{https://doi.org/10.1103/PhysRevX.8.021048}{Phys. Rev. X {\bf 8}, 021048 (2018)}.

\bibitem{Iskakov2024} S. Iskakov, M. I. Katsnelson,  A. I. Lichtenstein, Perturbative solution of fefrmionic sign problem in quantum Monte Carlo computations, \href{https://doi.org/10.1038/s41524-024-01221-w}{npj Computational Materials {\bf 10}, 36 (2024).}

\bibitem{Stepanov2026} E.A. Stepanov, S. Iskakov, M.I. Katsnelson, A.I. Lichtenstein, Superconductivity of Bad Fermions: Origin of Two Gaps in HTSC Cuprates, \href{https://doi.org/10.1038/s42005-026-02532-8}{Comm. Physics 9, {\bf 91} (2026)}. 


\bibitem{TremblayKyungSenechal} A.-M. S. Tremblay, B. Kyung, and D. S\'en\'echal, Pseudogap and high-temperature superconductivity from weak to strong coupling. Towards a quantitative theory, \href{https://doi.org/10.1063/1.2199446}{Low Temp. Phys. {\bf 32}, 424 (2006)}.

\bibitem{Gull} E. Gull, O. Parcollet, and A. J. Millis, Superconductivity and the Pseudogap in the two-dimensional Hubbard model, \href{https://doi.org/10.1103/PhysRevLett.110.216405}{Phys. Rev. Lett. {\bf 110}, 216405 (2013)}.

\bibitem{Gunnarson} O. Gunnarsson, T. Schäfer, J. P. F. LeBlanc, E. Gull, J. Merino, G. Sangiovanni, G. Rohringer, A. Toschi, Fluctuation diagnostics of the electron self-energy: Origin of the pseudogap physics, \href{https://doi.org/10.1103/PhysRevLett.114.236402}{Phys. Rev. Lett. {\bf 114}, 236402 (2015)}.

\bibitem{Krien2025} Y. Yu, S. Iskakov, E. Gull, K. Held, and F. Krien, Unambiguous Fluctuation Decomposition of the Self-Energy: Pseudogap Physics beyond Spin Fluctuations, \href{https://doi.org/10.1103/PhysRevLett.132.216501}{Phys. Rev. Lett. 132, 216501 (2024)}; Pairing boost from enhanced spin-fermion coupling in the pseudogap regime, Phys. Rev. B {\bf 112}, L041105 (2025).


\bibitem{Krien2022} F. Krien, P. Worm, P. Chalupa, A. Toschi, and K. Held, Explaining the pseudogap through damping and antidamping on the Fermi surface by imaginary spin scattering, \href{https://dx.doi.org/10.1038/s42005-022-01117-5}{Commun. Phys. {\bf 5}, 336 (2022).}

\bibitem{Kugler2025} J.-M. Lihm, D. Kiese, S.-S. B. Lee, F. B. Kugler, The finite-difference parquet method: Enhanced electron-paramagnon scattering opens a pseudogap, \href{https://doi.org/10.1073/pnas.2525308123}{PNAS {\bf 123}, e2525308123 (2026)}.

\bibitem{tJ1} E. Dagotto, Correlated electrons in high-temperature superconductors, \href{ https://doi.org/10.1103/RevModPhys.66.763}{Rev. Mod. Phys. {\bf 66}, 763 (1994).}

\bibitem{tJ2} M. Imada, A. Fujimori, and Y. Tokura, Metal-insulator transitions, \href{https://doi.org/10.1103/RevModPhys.70.1039}{Rev. Mod. Phys. {\bf 70}, 1039 (1998)}.

\bibitem{tJ3} P. A. Lee, N. Nagaosa, and X.-G. Wen, Doping a Mott insulator: Physics of high-temperature superconductivity, \href{https://doi.org/10.1103/RevModPhys.78.17}{Rev. Mod. Phys. {\bf 78}, 17 (2006)}.

\bibitem{Milstein2008} A.~I.~Milstein and O.~P.~Sushkov, \newblock Effective action, magnetic excitations, and quantum fluctuations in lightly doped single-layer cuprates, \newblock \href{https://doi.org/10.1103/PhysRevB.78.014501}{Phys. Rev. B \textbf{78}, 014501} (2008).




\bibitem{NikolaenkoSachdev2023} A. Nikolaenko, J. von Milczewski, D. G. Joshi, and S. Sachdev, Spin density wave, Fermi liquid, and fractionalized phases in a theory of antiferromagnetic metals using paramagnons and bosonic spinons, \href{https://doi.org/10.1103/PhysRevB.108.045123}{Phys. Rev. B {\bf 108}, 045123 (2023)}.


\bibitem{ZhangSachdev2020} Y.-H. Zhang and S. Sachdev, Deconfined criticality and ghost Fermi surfaces at the onset of antiferromagnetism in a metal, \href{https://doi.org/10.1103/PhysRevB.102.155124}{Phys. Rev. B {\bf 102}, 155124 (2020)}.

\bibitem{SachdevGaugeTheory2019} S. Sachdev, H. D. Scammell, M. S. Scheurer, and G. Tarnopolsky, Gauge theory for the cuprates near optimal doping, \href{https://doi.org/10.1103/PhysRevB.99.054516}{Phys. Rev. B {\bf 99}, 054516 (2019)}.

\bibitem{SachdevBergChatterjeeSchattner2016} S. Sachdev, E. Berg, S. Chatterjee, and Y. Schattner, Spin density wave order, topological order, and Fermi surface reconstruction, \href{https://doi.org/10.1103/PhysRevB.94.115147}{Phys. Rev. B {\bf 94}, 115147 (2016)}.

\bibitem{Sachdev} S. Sachdev, M. A. Metlitski, Y. Qi, and C. Xu, Fluctuating spin density waves in metals, \href{https://doi.org/10.1103/PhysRevB.80.155129}{Phys. Rev. B \textbf{80}, 155129 (2009)}.



\bibitem{QiSachdev2010} Y. Qi and S. Sachdev, Effective theory of Fermi pockets in fluctuating antiferromagnets, \href{https://doi.org/10.1103/PhysRevB.81.115129}{Phys. Rev. B {\bf 81}, 115129 (2010)}.

\bibitem{SachdevHiggs1} D. Chowdhury and S. Sachdev, Higgs criticality in a two-dimensional metal, \href{https://doi.org/10.1103/PhysRevB.91.115123}{Phys. Rev. B {\bf 91}, 115123 (2015)}.

\bibitem{ChatterjeeSachdevEberlein2017} S. Chatterjee, S. Sachdev, and A. Eberlein, Thermal and electrical transport in metals and superconductors across antiferromagnetic and topological quantum transitions, \href{https://doi.org/10.1103/PhysRevB.96.075103}{Phys. Rev. B {\bf 96}, 075103 (2017)}.

\bibitem{ChatterjeeSachdevScheurer2017} S. Chatterjee, S. Sachdev, and M. Scheurer, Intertwining topological order and broken symmetry in a theory of fluctuating spin density waves, \href{https://doi.org/10.1103/PhysRevLett.119.227002}{Phys. Rev. Lett. {\bf 119}, 227002 (2017)}.

\bibitem{StepanovBrenerHarkovKatsnelsonLichtenstein2022} E. A. Stepanov, S. Brener, V. Harkov, M. I. Katsnelson, and A. I. Lichtenstein, Spin dynamics of itinerant electrons: local magnetic moment formation and Berry phase, \href{https://doi.org/10.1103/PhysRevB.105.155151}{Phys. Rev. B {\bf 105}, 155151 (2022)}.

\bibitem{VilardiBonetti2025} D. Vilardi and P. M. Bonetti, SC$^*$ superconductivity and spin stiffnesses in the SU(2) gauge theory of the two-dimensional Hubbard model, \href{https://doi.org/10.48550/arXiv.2511.03436}{arXiv:2511.03436 [cond-mat.str-el] (2025)}.

\bibitem{MullerGroelingBonettiForniMetzner2026} H. M\"uller-Groeling, P. M. Bonetti, P. Forni, and W. Metzner, SU(2) gauge theory of fluctuating stripe order in the two-dimensional Hubbard model, \href{https://doi.org/10.48550/arXiv.2603.13071}{arXiv:2603.13071 [cond-mat.str-el] (2026)}.

\bibitem{BonettiNew} D. Vilardi, P. M. Bonetti, and W. Metzner, Spin stiffnesses and stability of magnetic order in the lightly doped two-dimensional Hubbard model, \href{https://doi.org/10.1103/x7qr-f6lm}{Phys. Rev. B {\bf 112}, 245149 (2025)}.

\bibitem{ForniBonettiMullerGroelingVilardiMetzner2026} P. Forni, P. M. Bonetti, H. M\"uller-Groeling, D. Vilardi, and W. Metzner, Spin susceptibility in a pseudogap state with fluctuating spiral magnetic order, \href{https://doi.org/10.1103/zm7b-jdzf}{Phys. Rev. B {\bf 113}, 045144 (2026)}.




\bibitem{BonettiMetzner} P. M. Bonetti and W. Metzner, SU(2) gauge theory of the pseudogap phase in the two-dimensional Hubbard model, \href{https://doi.org/10.1103/PhysRevB.106.205152}{Phys. Rev. B \textbf{106}, 205152 (2022)}.



\bibitem{OurSecond} I. A. Goremykin and A. A. Katanin, Antiferromagnetic and spin spiral correlations in the doped two-dimensional Hubbard model: gauge symmetry, Ward identities, and dynamical mean-field theory analysis, \href{https://dx.doi.org/10.1103/PhysRevB.110.085153}{Phys. Rev. B {\bf 110}, 085153 (2024)}



\bibitem{OurThird} I. A. Goremykin and A. A. Katanin, Frequency dependence of temporal spin stiffness and short-range magnetic order in the doped two-dimensional Hubbard model, \href{https://dx.doi.org/10.1103/w4vc-n5l6}{Phys. Rev. B {\bf 112}, L060405 (2025)}

\bibitem{DMFT_Incomm1} P. M. Bonetti, J. Mitscherling, D. Vilardi, and W. Metzner, Charge carrier drop at the onset of pseudogap behavior in the two-dimensional Hubbard model, \href{https://doi.org/10.1103/PhysRevB.101.165142}{Phys. Rev. B {\bf 101}, 165142 (2020)}.

\bibitem{OurFirst} I. A. Goremykin and A. A. Katanin, Commensurate and spiral magnetic order in the doped two-dimensional Hubbard model: Dynamical mean-field theory analysis, \href{https://doi.org/10.1103/PhysRevB.107.245104}{Phys. Rev. B \textbf{107}, 245104 (2023)}.





\bibitem{Schulz} H. J. Schulz, Effective action for strongly correlated fermions from functional integrals, \href{https://doi.org/10.1103/PhysRevLett.65.2462}{Phys. Rev. Lett. \textbf{65}, 2462 (1990)}; H. J. Schulz, Functional Integrals for Correlated Electrons, Proceedings of NATO Advanced Research Workshop on the Physics and Mathematical Physics of the Hubbard Model (Springer, New York, 1995), \href{https://doi.org/10.1007/978-1-4899-1042-4}{Vol. \textbf{343}, p. 89}. 

\bibitem{Weng} Z. Y. Weng, C. S. Ting, and T. K. Lee, Path-integral approach to the Hubbard model, \href{https://doi.org/10.1103/PhysRevB.43.3790}{Phys. Rev. B \textbf{43}, 3790 (1991)}.

\bibitem{Dupuis} K. Sengupta and N. Dupuis, Effective action and collective modes in quasi-one-dimensional spin-density-wave systems, \href{https://doi.org/10.1103/PhysRevB.61.13493}{Phys. Rev. B \textbf{61}, 13493 (2000)}; Y. Tomio, N. Dupuis, and Y. Suzumura, Effect of nearest- and next-nearest neighbor interactions on the spin-wave velocity of one-dimensional quarter-filled spin-density-wave conductors, \href{https://doi.org/10.1103/PhysRevB.64.125123}{Phys. Rev. B \textbf{64}, 125123 (2001)}.

\bibitem{Dupuis1} N. Dupuis, Spin fluctuations and pseudogap in the two-dimensional half-filled Hubbard model at weak coupling, \href{https://doi.org/10.1103/PhysRevB.65.245118}{Phys. Rev. B \textbf{65}, 245118 (2002)}; K. Borejsza and N. Dupuis, Antiferromagnetism and single-particle properties in the two-dimensional half-filled Hubbard model: A nonlinear sigma model approach, \href{https://doi.org/10.1103/PhysRevB.69.085119}{Phys. Rev. B \textbf{69}, 085119 (2004)}.

 \bibitem{DMFT_Incomm_Licht} M. Fleck, A. I. Liechtenstein, A. M. Ole\'s, L. Hedin, and V. I. Anisimov, Dynamical Mean-Field Theory for Doped Antiferromagnets, \href{https://dx.doi.org/10.1103/PhysRevLett.80.2393}{Phys. Rev. Lett. {\bf 80}, 2393 (1998)}.

\bibitem{DMFT_Incomm} S. Goto, S. Kurihara, and D. Yamamoto, Incommensurate spiral magnetic order on anisotropic triangular lattice: Dynamical mean-field study in a spin-rotating frame, \href{https://dx.doi.org/10.1103/PhysRevB.94.245145}{Phys. Rev. B {\bf 94}, 245145 (2016)}.


\bibitem{Luttinger} J.M. Luttinger, Fermi Surface and Some Simple Equilibrium Properties of a System of Interacting Fermions, \href{https://doi.org/10.1103/PhysRev.119.1153}{Phys. Rev. 119, 1153 (1960)}.

\bibitem{DzyaloshinskiiLuttinger} I. Dzyaloshinskii, Extended Van-Hove Singularity and Related Non-Fermi Liquids, \href{https://dx.doi.org/10.1051/jp1:1996127}{J. Phys. I France {\bf 6} 119 (1996)}; Some consequences of the Luttinger theorem: The Luttinger surfaces in non-Fermi liquids and Mott insulators, \href{https://doi.org/10.1103/PhysRevB.68.085113}{Phys. Rev. B {\bf 68}, 085113 (2003)}.

\bibitem{LuttingerSurface_Kitatani2025} M. Kitatani, Y. Nomura, S. Sakai, and R. Arita, Luttinger surface and exchange splitting induced by ferromagnetic fluctuations, \href{https://doi.org/10.48550/arXiv.2509.21034}{arXiv:2509.21034 (2025)}.

\bibitem{ToschiLuttinger} P. Worm, M. Reitner, K. Held, and A. Toschi, Fermi and Luttinger Arcs: Two Concepts, Realized on One Surface, \href{https://doi.org/10.1103/PhysRevLett.133.166501}{Phys. Rev. Lett. {\bf 133}, 166501 (2024)}.





























































\end{thebibliography}
\end{document}